\documentclass{article}

\usepackage{fullpage,times,comment,epsfig,algorithm,algorithmic,graphicx,cite,multirow,enumerate,amssymb,amsthm,url}

\usepackage[cmex10]{amsmath}
\usepackage{array}
\usepackage{mdwtab}
\usepackage[tight,footnotesize]{subfigure}
\usepackage{tikz}
\usepackage[para]{threeparttable}

\newcommand{\sknnb}{S$k$NN$_\textrm{b}$}
\newcommand{\sknnm}{S$k$NN$_\textrm{m}$}

\usepackage{tikz-qtree}
\usetikzlibrary{arrows}

\tikzset{
  treenode/.style = {align=left, inner sep=.5pt, text centered,
    font=\sffamily \tiny},
  arn_n/.style = {treenode, circle, white, font=\sffamily\bfseries, draw=black,
    fill=black, text width=1em},% arbre rouge noir, noeud noir
    arn_r/.style = {treenode, circle, red, draw=red, 
    text width=1em, very thick},% arbre rouge noir, noeud rouge
  arn_x/.style = {treenode, rectangle, draw=black,
    minimum width=0.3em, minimum height=0.3em}% arbre rouge noir, nil
}

\newtheorem{example}{Example}
\newtheorem{definition}{Definition}

\begin{document}
\begin{center}
\vspace*{5cm}
{\huge\bf\sf Secure {\em k}-Nearest Neighbor Query over\\ Encrypted Data in Outsourced Environments}\\
\vspace{3cm}

{\Large\bf\sf Yousef Elmehdwi, Bharath K. Samanthula and Wei Jiang}\\
%\emph{Department of Computer Science\\Missouri University of Science and Technology\\Rolla, Missouri, 65409, United States.}\\

{\large\sf \emph{Email:} \{ymez76, bspq8, wjiang\}@mst.edu}\\

\vspace{3cm}
{\Large\sf \today}\\ %adds the current date
\vspace{1cm}
\hrule
\begin{center}
{\large\sf Technical Report \\
Department of Computer Science, Missouri S\&T\\
500 West 15th Street, Rolla, Missouri 65409}\\
\end{center}

\begin{comment}
\begin{figure}[!h]
\centering
\epsfig{file=logo-new, width= .3\textwidth}
%\caption{Sample snapshot (two-hop) of social network for alex}
\label{fig: logo}
\end{figure}
\end{comment}
\end{center}

\title{}
%\title{Secure {\em k}-Nearest Neighbor Query over Encrypted Data}
\author{}

%\author{Yousef Elmehdwi, Bharath K. Samanthula and Wei Jiang\\
%Department of Computer Science, Missouri S\&T\\
%Missouri University of Science \& Technology\\
%500 West 15th Street, Rolla, Missouri 65409\\
%Email: \{ymez76,bspq8, wjiang\}@mst.edu}

\date{}
\maketitle

\begin{abstract}
For the past decade, query processing on relational data has been studied extensively, 
and many theoretical and practical  
solutions to query processing have been proposed under various scenarios. 
With the recent popularity of cloud computing, 
users now have the opportunity to outsource their data as well as the data management 
tasks to the cloud. However, due 
to the rise of various privacy issues, sensitive data (e.g., medical records) need to be
encrypted before outsourcing to the cloud. In addition, query processing tasks should be handled 
by the cloud; otherwise, there would be no point to outsource the data at the first place. 
To process queries over encrypted data without the cloud ever decrypting the data is a 
very challenging task.  
In this paper, 
we focus on solving the $k$-nearest neighbor ($k$NN) query problem over encrypted database
outsourced to a cloud: a user issues an encrypted query record to the cloud, and the 
cloud returns the $k$ closest records to the user. 
We first present a basic scheme  
and demonstrate that such a naive solution is not secure. To provide better security, 
we propose a secure $k$NN protocol that protects  
the confidentiality of the data, user's input query, and data access patterns. Also, we empirically analyze 
the efficiency of our protocols through various experiments. These results indicate that
our secure protocol is very efficient on the user end, and this 
lightweight scheme allows a user to use any mobile device to perform the $k$NN query.\\

\end{abstract}
%%%%%%%%%%%%%%% I N T R O D U C T I O N %%%%%%%%%%%%%%%%%%%%%%%%%%%%%
\section{Introduction}
\label{sec:intr}
As an emerging computing paradigm, cloud computing attracts many 
organizations to consider utilizing the benefits of a cloud in terms of  
cost-efficiency, flexibility, and offload of administrative overhead. 
In cloud computing model \cite{hu2011processing,mell2011nist}, a data 
owner outsources his/her database $T$ and the DBMS functionalities to the 
cloud that has the infrastructure to host outsourced databases and 
provides access mechanisms for querying and managing the hosted database. 
On one hand, by outsourcing, the data owner gets the benefit of reducing the data management
 costs and improves the quality of service. On the other hand, hosting and query processing of data out
 of the data owner control raises security challenges such as preserving data confidentiality and query privacy.

One straightforward way to protect the confidentiality of 
the outsourced data from the cloud as well 
as from the unauthorized users is to encrypt data by the data owner before outsourcing\cite{abadi2009data,pearson2009privacy,li2012toward}. 
By this way, the data owner can protect the privacy of his/her own data. 
In addition, to preserve query privacy, authorized users require encrypting their queries before sending them to the cloud for evaluation.
Furthermore, during query processing,
 the cloud can also derive useful and sensitive information 
about the actual data items by observing the data access 
patterns even if the data and query are encrypted \cite{williams2008building,de2012managing}.
Therefore, following from the above discussions, secure query processing needs to guarantee (1) 
confidentiality of the encrypted data (2) 
confidentiality of a user's query record and (3) hiding data access patterns.

Using encryption as a way to achieve data confidentiality may cause another issue
during the query processing step in the cloud. In general, it is very
difficult to process encrypted data without ever having to decrypt it. 
The question here is how the cloud can execute the queries over encrypted data while the data stored at the 
cloud are encrypted at all times.
In the literature, various techniques related to query processing over 
encrypted data have been proposed, including range queries \cite{agrawal2004order,hore2004privacy,shi2007multi, hore2012secure} 
and other aggregate queries \cite{hacigumucs2004efficient,mykletun2006aggregation}.  
However, these techniques are either not applicable or inefficient to solve advanced queries such 
as the $k$-nearest neighbor ($k$NN) query. 

In this paper, we address the problem of secure processing 
of $k$-nearest neighbor query over encrypted data (S$k$NN) in the cloud. 
Given a user's input query $Q$, the objective of the S$k$NN problem is to securely 
identify the $k$-nearest data tuples to $Q$ using the encrypted database of 
$T$ in the cloud, without allowing the cloud to learn anything regarding the actual contents of 
the database $T$ and the query record $Q$.
More specifically, when encrypted data are outsourced to the cloud, we observe that 
an effective S$k$NN protocol needs to satisfy the following properties: 
\begin{itemize}\itemsep=0pt
\item Preserve the confidentiality of $T$ and $Q$ at all times
\item Hiding data access patterns from the cloud
\item Accurately compute the $k$-nearest neighbors of query $Q$
\item Incur low computation overhead on the end-user
 \end{itemize}
In the past few years, researchers have proposed various methods \cite{wong2009secure,hu2011processing,yaosecure} to 
address the S$k$NN problem. However, we emphasize that the 
existing S$k$NN methods proposed in \cite{wong2009secure,hu2011processing} violate at 
least one of the above mentioned desirable properties of a S$k$NN protocol.
On one hand, the methods in \cite{wong2009secure,hu2011processing} are insecure because they 
are vulnerable to chosen and known plaintext attacks. 
%For more details, we refer the reader to\cite{yaosecure}. 
On the other hand, recent method in \cite{yaosecure} returns non-accurate 
$k$NN result to the end-user. More precisely, in \cite{yaosecure}, the cloud retrieves the relevant encrypted 
partition instead of finding the encrypted exact $k$-nearest neighbors. Furthermore, in 
\cite{hu2011processing,yaosecure}, the end-user involves in heavy computations during the query processing step. 
By doing so, the method in \cite{yaosecure} utilizes cloud as just a storage medium, i.e., no significant 
work is done on the cloud side.   
More details about the existing S$k$NN methods are provided in Section \ref{sec:related-work}. 

Along this direction, with the goal of providing better security, this paper proposes a 
novel S$k$NN protocol that satisfies the above properties altogether. 
%%%%%%%%%%%%%%% P r o b l e m   D e f i n i t i o n %%%%%%%%%%%%%%%%%%%%%%%%%%%%%
\subsection{Problem Definition}\label{sec:problemfor}
Suppose the data owner Alice owns a database $T$ of 
$n$ records, denoted by $t_1, \ldots, t_n$,  
and $m$ attributes. Let
$t_{i,j}$ denote the $j^{th}$ attribute value of record $t_i$. In our problem setting, we 
assume that Alice 
initially encrypts her database attribute-wise, that is, 
she computes $E_{pk}(t_{i,j})$, for $1 \le i \le n$ and $1 \le j \le m$, where $E_{pk}$ denotes the encryption 
function of a public-key cryptosystem that is semantically 
secure \cite{paillier-99}. Let the encrypted database be denoted by $E_{pk}(T)$. We assume 
that Alice outsources $E_{pk}(T)$ as well as the future querying processing services to the cloud.

Consider an authorized user Bob who wants 
to ask the cloud for $k$-neighbor records that are closest to his 
input query $Q = \langle q_1, \ldots, q_m\rangle$ based on $E_{pk}(T)$. 
During this process, Bob's query $Q$ and contents of database $T$ should not be 
revealed to the cloud. In addition, the access patterns to the data should be protected 
from the cloud. We refer 
to such a process as Secure $k$NN (S$k$NN) query over encrypted 
data in the cloud. Without loss of generality, let $\langle t'_1, \ldots, t'_k \rangle$ denote 
the $k$-nearest records to $Q$. Then, we formally define the S$k$NN protocol as follows: 
$$\textrm{S}k\textrm{NN}(E_{pk}(T), Q) \rightarrow \langle t'_1,\dots,t'_k \rangle$$ 
We emphasize that, at the end of the S$k$NN protocol, the output $\langle t'_1,\dots,t'_k \rangle$ should 
be revealed only to Bob. We now present a real-life application of the S$k$NN protocol. 
%without revealing the plaintext of the query $Q$ and the database $T$ to the cloud.
%The cloud computes $k$ encrypted records closest to the query $Q$ 
%(say $E_{pk}(t'_1),\dots,E_{pk}(t'_k)$ and then sends $E_{pk}(t'_1),\dots,E_{pk}(t'_k)$ 
%to Bob. Once Bop receives $E_{pk}(t'_1),\dots,E_{pk}(t'_k)$, he decrypts 
%them and then obtains $t'_1,\dots,t'_k$ as the result of his query $Q$.
%We refer to such a process as a Secure $k$NN (S$k$NN) query over encrypted data in the cloud. 
%Formally, we define the S$k$NN protocol as:
%$$\textrm{S}k\textrm{NN}(E_{pk}(T), Q) \rightarrow t'_1,\dots,t'_k $$ 
%where $t'_1,\dots,t'_k$ denotes the set of $k$ records closest to the query $Q$ returned to Bob.
%%%%%%%%%%%%%%% E X A M P L E %%%%%%%%%%%%%%%%%%%%%%%

\renewcommand{\tabcolsep}{.15cm}
\begin{table}[!t]
\centering
%{\normalsize
\caption{Sample Heart Disease Dataset $T$} % title of Table
%\centering % used for centering table
\begin{tabular}{ccccccccccc}  
     \hline\hline %inserts double horizontal lines
     record-id &\;age&\;sex&\;cp&\;trestbps&\;chol&\;fbs&\;slope&\;ca&\;thal&\;num \\ [1ex] % inserts table
     \hline % inserts single horizontal line
    	$t_1$ &63 & 1 & 1 & 145 & 233 & 1 & 3 & 0 &6 & 0  \\ % inserting body of the table
           $t_2$ & 56 & 1  & 3	&130& 256 & 1& 2 & 1 & 6&2\\
    	$t_3$ & 57 & 0 & 3  & 140 & 241 & 0 & 2 & 0 & 7 & 1 \\
    	$t_4$ & 59 & 1 & 4 & 144 & 200 & 1 & 2 & 2 & 6 & 3 \\
    	$t_5$ & 55 & 0 & 4  & 128 & 205 & 0 & 2 & 1 &7 & 3 \\
    	$t_6$ & 77 & 1 & 4 & 125 & 304 & 0 & 1 & 3 & 3 & 4 \\ [1ex] % [1ex] adds vertical space
    \hline %inserts single line
\end{tabular}%}
\label{table:example} % is used to refer this table in the text
\end{table} 
%}

\begin{example}\label{sec:example}
%$k$NN query can be used as a primitive tool to get a possible diagnosis among a set of 
%possible diseases that have similar symptoms. For example, assume a
Consider a physician who would like to know the risk factor of heart disease in a specific patient.
Let $T$ denote the sample heart disease dataset with attributes \textit{record-id}, \textit{age}, \textit{sex}, \textit{cp}, 
\textit{trestbps}, \textit{chol}, \textit{fbs}, \textit{slope}, \textit{ca}, \textit{thal}, and \textit{num} 
as shown in Table \ref{table:example}. The description and range for each of these attributes 
are shown in Table \ref{table:map}. The heart disease dataset 
given in Table  \ref{table:example} is obtained from the UCI machine learning repository \cite{uci-dataset-heart}.

Initially, the data owner (hospital) encrypts $T$ attribute-wise, outsources the encrypted database $E_{pk}(T)$ 
to the cloud for easy management. In addition, the data owner delegates the future query processing 
services to the cloud. Now, we consider a doctor working at the hospital, say Bob, who would 
like to know the risk factor of heart disease in a specific 
patient based on $T$. Let the patient medical information 
be $Q = \langle 58,1, 4, 133, 196, 1, 2,1,$ $6\rangle$. In the S$k$NN protocol, Bob first need to encrypt 
$Q$ (to preserve the privacy of his query) and send it to the cloud. Then the cloud searches on the encrypted database 
$E_{pk}(T)$ to figure out the $k$-nearest neighbors to the user's request. For simplicity, 
let us assume $k=2$. Under this case, the $2$ nearest neighbors to $Q$ are $t_4$ and 
$t_5$ (by using Euclidean distance as the similarity metric). After this, the cloud 
sends $t_4$ and $t_5$ (in encrypted form) to Bob. Here, the cloud should identify the 
nearest neighbors of $Q$ in an oblivious manner without knowing any sensitive information, i.e., all 
the computations have to be carried over encrypted records. Finally, Bob receives  
%response from the cloud, he can decrypt it to obtain the information for his input query record $Q$ 
$t_4$ and $t_5$ that will help him to make medical decisions. 
\hfill $\Box$
\end{example}

\renewcommand{\tabcolsep}{.15cm}
\begin{table}[t]
%\footnotesize
\caption{Attribute Description of Heart Disease Dataset $T$}
\centering
\renewcommand{\arraystretch}{1.3}
\begin{tabular}{|l | l|}
\hline
\; age& \; age in years \\
\hline
\; sex&\; 1=male, 0=female \\ 
\hline
\; cp &\; chest pain type: 1=typical angina, 2=atypical angina,\\
&\;  3=non-anginal pain, 4=asymptomatic \\
\hline
 \; trestbps&\; resting blood pressure (mm Hg)\\
\hline
\; chol&\; serum cholesterol in mg/dl\\
\hline
 \; fbs&\; fasting blood sugar $>$ 120 mg/dl (1=true; 0=false) \\
\hline
\; slope&\; slope of the peak exercise ST segment \\
         &\; (1=upsloping, 2=flat, 3=downsloping) \\
\hline
 \; ca&\; number of major vessels (0-3) colored by flourosopy \\
\hline
 \; thal&\; 3=normal, 6=fixed defect, 7=reversible defect\\
\hline
 \; num&\; diagnosis of heart disease  from 0 (no presence) to 4\\
\hline 
\end{tabular}
\label{table:map}
\end{table}

%%%%%%%%%%%%%%% O u r   C o n t r i b u t i o n  %%%%%%%%%%%%%%%%%%%%%%%%%%%%%
\subsection{Our Contribution} 
In this paper, we propose a novel S$k$NN protocol to facilitate the $k$-nearest neighbor 
search over encrypted data in the cloud that preserves both the data privacy and query privacy. In our 
protocol, once the encrypted data are outsourced to the cloud, Alice does not participate in any 
computations. Therefore, no information is revealed to Alice. 
In particular, the proposed protocol meets the following requirements:
\begin{itemize}\itemsep=0pt
\item \textbf{Data confidentiality - } Contents of $T$ or any intermediate results should not be revealed to the cloud. 
\item \textbf{Query privacy - } Bob's input query $Q$ should not be revealed to the cloud.
\item  \textbf{Correctness - } The output $\langle t'_1,\dots,t'_k \rangle$ 
should be revealed only to Bob. In addition, no information other than $ t'_1,\dots,t'_k$ 
should be revealed to Bob.
\item \textbf{Low computation overhead on Bob - } After sending his encrypted 
query record to the cloud, Bob involves only in a little computation 
compared with the existing works \cite{wong2009secure,hu2011processing,yaosecure}. More details are 
given in Section \ref{sec:method}.
\item \textbf{Hidden data access patterns - } Access patterns to the data, such as 
the records corresponding to the $k$-nearest neighbors 
of $Q$, should 
not be revealed to Alice and the cloud (to prevent any inference attacks). 
\end{itemize}
We emphasize that the intermediate results seen by the cloud in our protocol 
are either newly generated randomized encryptions or 
random numbers. Thus, which data records 
correspond to the $k$-nearest neighbors of $Q$ are 
not known to the cloud.  In addition, after sending his encrypted 
query record to the cloud, Bob does not involve in any computations (less workload at Bob's local machine). Hence, 
data access patterns are further protected from Bob. More details are given in Section \ref{sec:method}.

The rest of the paper is organized as follows. 
We discuss the existing related work and some background concepts    
in Section \ref{sec:related-work}. A set of security primitives that are utilized 
in the proposed protocols and their possible implementations are provided in Section \ref{sec:sub-methods}. 
The proposed protocols are explained in detail in 
Section \ref{sec:proposed}. Section \ref{sec:exp} discusses the performance 
of the proposed protocols based on various experiments. We conclude the paper along 
with future work in Section \ref{sec:concl}.

%%%%%%%%%%%%%%% R e l a t e d   W o r k  a n d   B a c k g r o u n d %%%%%%%%%%%%%%%%%%%%%%%%%%%%%
\section{Related Work and Background}\label{sec:related-work}

In this section, we first present an overview of the existing secure $k$-nearest neighbor techniques. Then, we present 
the security definition adopted in this paper and the Paillier cryptosystem along with its additive 
homomorphic properties as a background. 
%%%%%%%%%%%%%%% S E C U R E  k N N TECHNIQUES %%%%%%%%%%%%%%%%%%%%%%%%%%%%%
\subsection{Existing S$k$NN Techniques}
%The goal of S$k$NN query is to protect the data using encryption and allowing encrypted query processing over the encrypted data without revealing the plain data to the server. 
Retrieving the  $k$-nearest neighbors to a given query $Q$ is one of the most 
fundamental problem in many application domains such as similarity search, 
pattern recognition, and data mining. In the literature, many techniques 
have been proposed to address the S$k$NN problem, which can be classified into 
two categories based on whether the data are encrypted or not: \textit{centralized} and \textit{distributed}.
%%%%%%%%%%%%%%% C e n t r a l i z e d %%%%%%%%%%%%%%%%%%%%%%%%%%%%%
\subsubsection{Centralized Methods}
In the centralized methods, we assume that the data owner outsources his/her database and DBMS 
functionalities (e.g., $k$NN query) to an untrusted external service provider which 
manages the data on behalf of the data owner where only trusted users 
are allowed to query the hosted data at the service provider. 
By outsourcing data to an untrusted server, many security issues arise, 
such as data privacy (protecting the confidentiality of the data from 
the server as well as from query issuer). 
% and query privacy (protect the content of user’s query from data owner and the server). 
To achieve data privacy, data owner is required to use data anonymization models 
(e.g., $k$-anonymity) or cryptographic (e.g., encryption and data perturbation) techniques 
over his/her data before outsourcing them to the server.
% Apart from the information loss, the data anonymization techniques, such as \cite{} suffer from ``disclosing the data or the query in a coarser and imprecise form''\cite{hu2011processing}.

Encryption is a traditional technique used to protect the confidentiality 
of sensitive data such as medical records. Due to data encryption, the process of 
query evaluation over encrypted data becomes challenging. 
Along this direction, various techniques have been proposed for processing range \cite{agrawal2004order,hore2004privacy,shi2007multi, hore2012secure} 
and aggregation queries \cite{hacigumucs2004efficient,mykletun2006aggregation} over encrypted data. 
%Range queries can be achieved using order preserving encryption scheme (OPES) that proposed in\cite{agrawal2004order}, 
%whereas \textit{aggregation queries}\cite{ge2007answering} achieved by applying homomorphic 
%encryption (additive/multiplicative)\cite{Paillier99}. 
However, in this paper, 
we restrict our discussion to secure evaluation of $k$NN query.

In the past few years, researchers have proposed different methods 
\cite{wong2009secure,hu2011processing,yaosecure} to  address the S$k$NN problem. 
Wong et al.\cite{wong2009secure} proposed a new encryption scheme called asymmetric 
scalar-product-preserving encryption (ASPE) that preserves scalar product  
between the query vector $Q$ and any tuple vector $t_i$ from database $T$ for 
distance comparison which is sufficient to find $k$NN. 
In\cite{wong2009secure}, data and query are encrypted using 
slightly different encryption schemes before outsourcing to the server.
As an alternative, Hu et al.\cite{hu2011processing} proposed a method based on Privacy 
Homomorphism (PH) encryption scheme. More specifically, they used a provably 
secure privacy homomorphism encryption scheme from\cite{domingo2002provably} 
that supports modular addition, subtraction and multiplication over encrypted data. They addressed 
the S$k$NN problem under the following setting: the client has the ciphertexts of all data 
points in database $T$ and the encryption function of $T$ whereas the server has 
the decryption function of $T$ and some auxiliary information regarding each data 
point. However, both methods in \cite{wong2009secure,hu2011processing} are not secure because they are 
vulnerable to chosen-plaintext attacks. We refer the reader to\cite{yaosecure} for more details 
on these security issues.

Recently, Yao et al.\cite{yaosecure} designed a new S$k$NN method based on partition-based 
secure Voronoi diagram (SVD). Instead of asking the cloud to retrieve the exact $k$NN, 
they required, from the cloud, to retrieve a relevant encrypted partition $E_{pk}(G)$ for 
$E_{pk}(T)$ such that $G$ is guaranteed to contain the $k$-nearest neighbors of $Q$. 
However, in our work, we are able to solve the S$k$NN problem accurately by letting the  
cloud to retrieve the exact $k$-nearest neighbors of $Q$ (in encrypted form). 
In addition, most of the computations during the query 
processing step in \cite{hu2011processing,yaosecure} are performed locally by the end-user (i.e., query issuer)  
which conflicts the very purpose of outsourcing the DBMS functionalities to the cloud. 
Since our proposed protocol solves the problem of finding $k$-nearest neighbors 
of $E_{pk}(Q)$ over encrypted data, 
it can also be used in other relevant data mining tasks 
%that required ensuring of privacy preserving over encrypted data. 
such as secure clustering, classification, and outlier detection. 

%%%%%%%%%%%%%%% D i s t r i b u t e d  %%%%%%%%%%%%%%%%%%%%%%%%%%%%%  
\subsubsection{Data Distribution Methods}
In the data distributed methods, data are assumed to be partitioned either 
vertically or horizontally and distributed among a set of independent, 
non-colluding parties. 
%Under such an architecture, the parties want to conduct a computation based on their private inputs such
%that they learn the outputs without revealing their own inputs
%or outputs to others, and no party can infer anything
%other than what can be learned from its own input and output \cite{qi2008efficient}. 
In the literature, the data distributed methods 
rely on secure multiparty computation (SMC) techniques 
that enable multiple parties to securely evaluate a function using their respective 
private inputs without disclosing the input of one party to the others. Many 
efforts have been made to address the problem of $k$NN query in a distributed 
environment. Shaneck et al.\cite{shaneck2009privacy} proposed privacy-preserving algorithm to perform  
$k$-nearest neighbor search.
%, were the first to introduce a protocol for Privacy-Preserving Nearest Neighbor Search (PPNNS). 
The protocol in\cite{shaneck2009privacy} is based on secure multiparty computation for privately 
computing $k$NN points in a horizontally partitioned dataset. 
Qi et al.\cite{qi2008efficient} proposed a single-step $k$NN search protocol that 
is provably secure with linear computation and communication complexities. 
%extended the work in\cite{shaneck2009privacy} 
%and proposed an efficient privacy-preserving $k$NN search protocol where the data are distributed between two parties.   
Vaidya et al.\cite{vaidya2005privacy} studied privacy-preserving top-$k$ queries in which the data are vertically partitioned. 
Ghinita et al.\cite{ghinita2008private} proposed a private information retrieval (PIR) framework 
for answering $k$NN queries in location-based services. However, 
their solution protects only the query privacy, i.e., it does not address data confidentiality and access 
pattern issues. 

We emphasize that the above data distribution methods are not 
applicable to perform $k$NN queries over encrypted data for two reasons: (1). 
In our work, we deal with encrypted form of database and query 
which is not the case in the above methods (2). The 
database in our case is encrypted and stored on the cloud 
whereas in the above methods it is partitioned (in plaintext format) 
among different parties.
%%%%%%%%%%%%%%% S e c u r i t y   D e f i n i t i o n  %%%%%%%%%%%%%%%%%%%%%%%%%%%%%  
\subsection{Security Definition}
In this paper, 
privacy/security is closely related to the 
amount of information disclosed during the execution of a protocol. 
There are many ways to define information disclosure. To maximize privacy or 
minimize information disclosure, we adopt the security definitions in the literature 
of secure multiparty computation (SMC) first introduced by Yao's Millionaires' problem
for which a provably secure
solution was developed \cite{Yao82,Yao86}.
In this paper, we assume that parties are
semi-honest; that is,
a semi-honest party (also referred to as honest-but-curious)
follows the rules of the protocol using its correct input,
but is free to later use what it sees during execution of the protocol
to compromise security. We refer the reader to \cite{smc-2004,Goldreichnc} 
for detailed security definitions and models. Briefly, the following 
definition captures the above discussion
regarding a secure protocol under the semi-honest model.

\begin{definition}\label{def:semi-honest}
Let $a_i$ be the input of party $P_i$, $\prod_i(\pi)$ be  $P_i$'s 
execution image of the protocol $\pi$ and $b_i$ be the result computed from $\pi$ for $P_i$.
$\pi$ is secure if $\prod_i(\pi)$ can be simulated from
%$<T_i, \prod_i(f), s>$ and %!!! I think this is wrong - too easy...
$\langle a_i, b_i \rangle$ and
distribution of the simulated image is computationally indistinguishable
from $\prod_i(\pi)$.

\end{definition}

%%%%%%%%%%%%%%% P a i l l i e r  C r y p t o s y s t e m  %%%%%%%%%%%%%%%%%%%%%%%%%%%%%  
\subsection{Paillier Cryptosystem}
The Paillier cryptosystem is an additive homomorphic and probabilistic 
asymmetric encryption scheme \cite{paillier-99}. Let $E_{pk}$ be the encryption 
function with public key $pk$ given by ($N, g$), where $N$ is a product of two large primes and 
$g$ is in $\mathbb{Z}_{N^2}^*$. Also, let $D_{sk}$ be 
the decryption function with secret key $sk$. 
Given $a, b~\in~\mathbb{Z}_N$, the Paillier encryption scheme exhibits the following properties:

\begin{enumerate}[a.]
     \item \textbf{Homomorphic Addition -} $E_{pk}(a + b) \gets E_{pk}(a) \ast E_{pk}(b) \bmod N^2;$
     \item \textbf{Homomorphic Multiplication -} $E_{pk}(a\ast b) \gets E_{pk}(a)^{b} \bmod N^2;$
     \item \textbf{Semantic Security -} The encryption scheme is semantically 
secure\cite{goldwasser-89,Goldreichnc}. Briefly, given a set of ciphertexts, an 
adversary cannot deduce any additional information about the plaintext. 
\end{enumerate}
In this paper, we assume that a data owner encrypted his or her data using 
Paillier cryptosystem before outsourcing them to a cloud. Some common 
notations that are used extensively in this paper are shown in Table \ref{tb:notations}.
%%%%%%%%%%%%%%% P r i v a c y-P r e s e r v i n g  P r i m i t i v e s  %%%%%%%%%%%%%%%%%%%%%%%%%%%%%   
\section{ Basic Security Primitives}\label{sec:sub-methods}
In this section, we present a set of generic protocols that will be used 
as sub-routines while constructing our proposed S$k$NN protocol in Section \ref{sec:method}. All of 
the below protocols are considered under two-party semi-honest setting. In particular, 
we assume the existence of two semi-honest parties $P_1$ and $P_2$ such that the 
Paillier's secret key $sk$ is known only to $P_2$ whereas $pk$ is treated as public. 
\begin{table}[t]
%\footnotesize
\caption{Common Notations}
\centering
\renewcommand{\arraystretch}{1.3}
\begin{tabular}{| l | l |}
\hline
~Alice & ~The data owner of database $T$\\
\hline
~$E_{pk}(T)$ & ~Attribute-wise encryption of $T$ \\ 
\hline
~Bob & ~An authorized user who can access $E_{pk}(T)$ in the cloud \\
\hline
~$n$ & ~Number of data records in $T$\\
\hline
~$m$ & ~Number of attributes in $T$ \\
\hline
~$t_i$ & ~$i^{th}$ record in $T$ \\
\hline
~$Q$ & ~Bob's query record \\
\hline
~$t'_i$ & ~$i^{th}$ nearest record to $Q$ based on $T$\\
\hline
~$l$ & ~Domain size (in bits) of the squared Euclidean distance\\
& ~based on $T$ \\
\hline
~$\langle z_1, z_{l}\rangle$ & ~The most and least significant bits of integer $z$\\
\hline
~$[z]$ & ~Vector of encryptions of the individual bits of $z$\\
\hline 
\end{tabular}
\label{tb:notations}
\end{table}
  
\begin{itemize}
%%%%%%%%%%%%%%%     S M     %%%%%%%%%%%%%%%%%%%%%%%%%%%%   
\item Secure Multiplication (SM) Protocol:\\ 
This protocol considers $P_1$ 
with input $(E_{pk}(a), E_{pk}(b))$ and outputs $E_{pk}(a\ast b)$ to $P_1$, where $a$ and 
$b$ are not known to $P_1$ and $P_2$. 
During this process, no information regarding $a$ and $b$ is revealed to $P_1$ and $P_2$. The output $E_{pk}(a\ast b)$ is known only to $P_1$.
%%%%%%%%%%%%%%% S S E D  %%%%%%%%%%%%%%%%%%%%%%%%%%%%%    
\item Secure Squared Euclidean Distance (SSED) Protocol:\\ 
$P_1$ with input $(E_{pk}(X), E_{pk}(Y))$ 
and $P_2$ securely compute the encryption of squared Euclidean distance between vectors $X$ and $Y$. 
Here $X$ and $Y$ are $m$ dimensional vectors where $E_{pk}(X) = \langle E_{pk}(x_1), \ldots, E_{pk}(x_m)\rangle $ and 
$E_{pk}(Y) = \langle E_{pk}(y_1), \ldots, E_{pk}(y_m)\rangle$. At the end, the output  
$E_{pk}(|X - Y|^2)$ is known only to $P_1$.
%%%%%%%%%%%%%%% S B D   %%%%%%%%%%%%%%%%%%%%%%%%%%%%%   
\item Secure Bit-Decomposition (SBD) Protocol:\\ 
$P_1$ with input $E_{pk}(z)$ and $P_2$ securely compute the 
encryptions of the individual bits of $z$, where $0 \le z < 2^l$. The 
output $[z] = \langle E_{pk}(z_1), \ldots, E_{pk}(z_l)\rangle $ is known only to $P_1$. Here $z_1$ and $z_l$ 
denote the most and least significant bits of integer $z$ respectively.
%%%%%%%%%%%%%%% S M I N  %%%%%%%%%%%%%%%%%%%%%%%%%%%%%   
\item Secure Minimum (SMIN) Protocol:\\ 
$P_1$ with input $([u], [v])$ and $P_2$ with $sk$ securely compute 
the encryptions of the individual bits of minimum number between $u$ and $v$. 
That is, the output is $[\min(u, v)]$ which will be known only to $P_1$. During this protocol, 
no information regarding $u$ and $v$ is revealed to $P_1$ and $P_2$.
%%%%%%%%%%%%%%% S M I N_k   %%%%%%%%%%%%%%%%%%%%%%%%%%%%%   
\item Secure Minimum out of $n$ Numbers (SMIN$_n$) Protocol:\\ 
In this protocol, $P_1$ has $n$ encrypted vectors $([d_1], \ldots, [d_n])$ and 
$P_2$ has $sk$. Here $[d_i] = \langle E_{pk}(d_{i,1}), \ldots, E_{pk}(d_{i,l}) \rangle$ such that 
$d_{i,1}$ and $d_{i,l}$ are the most and least significant bits of integer $d_{i}$ respectively, 
for $1 \le i \le n$. $P_1$ and $P_2$ jointly compute the output $[\min(d_1,\ldots, d_n)]$. 
At the end of this protocol, $[\min(d_1,\ldots, d_n)]$ is known only to $P_1$. During the SMIN$_n$ 
protocol, no information regarding any of $d_i$'s is revealed to $P_1$ and $P_2$.
%%%%%%%%%%%%%%% S B O R   %%%%%%%%%%%%%%%%%%%%%%%%%%%%%    
\item Secure Bit-OR (SBOR) Protocol:\\
$P_1$ with input $(E_{pk}(o_1), E_{pk}(o_2))$ and $P_2$ securely compute 
$E_{pk}(o_1\vee o_2)$, where $o_1$ and $o_2$ are two bits. The output $E_{pk}(o_1\vee o_2)$ is 
known only to $P_1$. 
\end{itemize}
We now discuss each of these protocols in detail. 
Also, we either propose new solution or refer to the most efficient 
known implementation to each one of them. \\
%Also, please note that we may find similar named protocols 
%in the literature of Secure Multiparty Compuation. However, to our knowledge,
%we have yet to discover similar protocols that work on semantically secure 
%encrypted data. 

%%%%%%%%%%%%%%% Secure Multiplication (SM)   %%%%%%%%%%%%%%%%%%%%%%%%%%%%%   
\noindent \textbf {Secure Multiplication (SM). }
Consider a party $P_1$ with private input $(E_{pk}(a), E_{pk}(b))$ and a 
party $P_2$ with the secret key $sk$. The goal of the secure multiplication (SM) 
protocol is to return the encryption 
of $a \ast b$, i.e., $E_{pk}(a*b)$ as output to $P_1$. During this protocol, no information regarding 
$a$ and $b$ is revealed to $P_1$ and $P_2$. The basic idea of the SM protocol 
is based on the following property which holds 
for any given $a,b \in \mathbb{Z}_N$: 
\begin{equation}\label{eq:mult}
 a\ast b = (a+r_a)\ast (b+r_b) - a\ast r_b - b\ast r_a - r_a\ast r_b
\end{equation}
\begin{algorithm}[t]
\begin{algorithmic}[1]
\REQUIRE $P_1$ has $E_{pk}(a)$ and $E_{pk}(b)$; $P_2$ has $sk$
\STATE $P_1$:
\begin{enumerate}\itemsep=0pt
    \item[(a).]  Pick two random numbers $r_a, r_b \in \mathbb{Z}_N$
    \item[(b).]  $a' \gets E_{pk}(a)\ast E_{pk}(r_a)$
    \item[(c).]  $b' \gets E_{pk}(b)\ast E_{pk}(r_b)$; send $a', b'$ to $P_2$    
\end{enumerate}
\STATE $P_2$:
\begin{enumerate}\itemsep=0pt
    \item[(a).]  Receive $a'$ and $b'$ from $P_1$ 
    \item[(b).]  $h_a \gets D_{sk}(a')$;~ $h_b \gets D_{sk}(b')$
    \item[(c).] $h \gets h_a \ast h_b \bmod N$
    \item[(d).] $h' \gets E_{pk}(h)$; send $h'$ to $P_1$
\end{enumerate}
\STATE $P_1$:
\begin{enumerate}\itemsep=0pt
    \item[(a).]  Receive $h'$ from $P_2$ 
    \item[(b).]  $s \gets h' \ast E_{pk}(a)^{N- r_b}$
    \item[(c).]  $s' \gets s \ast E_{pk}(b)^{N- r_a}$
    \item[(d).]  $E_{pk}(a\ast b) \gets s'\ast E_{pk}(r_a\ast r_b)^{N-1}$
\end{enumerate}
\end{algorithmic}
\caption{SM$(E_{pk}(a), E_{pk}(b)) \rightarrow E_{pk}(a\ast b)$}
\label{alg:sm}
\end{algorithm}
\noindent where all the arithmetic operations are performed under $\mathbb{Z}_N$. The overall 
steps in SM are shown in Algorithm \ref{alg:sm}. Briefly, $P_1$ initially 
randomizes $a$ and $b$ by computing $a' = E_{pk}(a)*E_{pk}(r_a)$ and $b' = E_{pk}(b)*E_{pk}(r_b)$, and 
sends them to $P_2$. Here $r_a$ and $r_b$ are random numbers in $\mathbb{Z}_N$ known only to $P_1$. 
Upon receiving, $P_2$ decrypts and multiplies them to get $h = (a+r_a)\ast(b+r_b) \bmod N$. 
Then, $P_2$ encrypts $h$ and sends it to $P_1$. After this, $P_1$ removes extra random factors 
from $h' = E_{pk}((a+r_a)*(b+r_b))$ based on Equation \ref{eq:mult} to get $E_{pk}(a*b)$. 
Note that, for any given $x\in \mathbb{Z_N}$,  ``$N-x$'' is equivalent to ``$-x$'' under $\mathbb{Z}_N$. 
Hereafter, we use the notation $r \in_R \mathbb{Z}_N$ to denote $r$ as a random number in $\mathbb{Z}_N$.
\begin{example} Suppose $a = 59$ and $b = 58$. For simplicity, let $r_a = 1$ and $r_b = 3$. Initially, 
$P_1$ computes $a'= E_{pk}(60) = E_{pk}(a)*E_{pk}(r_a)$, $b' =E_{pk}(61) = E_{pk}(b)*E_{pk}(r_b)$ and 
sends them to $P_2$. Then, $P_2$ decrypts  
%$60\gets D_{sk}(E_{pk}(60))$ and $61\gets D_{sk}(E_{pk}(61))$) 
and multiplies them to get $h= 3660$. After this, $P_2$ encrypts $h$ to get $h'= E_{pk}(3660)$ and sends it to $P_1$. 
Upon receiving $h'$, $P_1$ computes $s = E_{pk}(3483)  = E_{pk}(3660 - a \ast r_b)$, and 
$s'= E_{pk}(3425) = E_{pk}(3483 - b \ast r_a)$. Finally, $P_1$ computes 
$E_{pk}(a \ast b) = E_{pk}(3422) = E_{pk}(3425 - r_a \ast r_b)$.
\hfill $\Box$\\
\end{example}
%%%%%%%%%%%%%%%%%%%%%%%%%%%% Squared Euclidean Distance  %%%%%%%%%%%%%%%%%%%%
\noindent \textbf{Secure Squared Euclidean Distance (SSED). }
In the SSED protocol, $P_1$ holds two encrypted vectors $(E_{pk}(X), E_{pk}(Y))$ and 
$P_2$ holds the secret key $sk$. Here $X$ and $Y$ are two $m$-dimensional vectors 
where $E_{pk}(X) = \langle E_{pk}(x_1), \ldots, E_{pk}(x_m)\rangle$ and 
$E_{pk}(Y) = \langle E_{pk}(y_1),\ldots, E_{pk}(y_m)\rangle$. The goal 
of the SSED protocol is to securely compute $E_{pk}(|X-Y|^2)$, where $|X-Y|$ denotes 
the Euclidean distance between vectors $X$ and $Y$. During this protocol, no information regarding 
$X$ and $Y$ is revealed to $P_1$ and $P_2$. 
The basic idea of SSED follows from 
following equation:
\begin{equation}\label{eq:euclidean}
|X-Y|^2 = \sum_{i=1}^m (x_i - y_i)^2 
\end{equation}
The main steps involved in SSED are shown in  Algorithm \ref{alg:ssed}. Briefly, for $1 \le i \le m$, $P_1$ initially 
computes $E_{pk}(x_i-y_i)$  by using the homomorphic properties. Then $P_1$ and $P_2$ jointly compute 
$E_{pk}((x_i-y_i)^2)$ using the SM protocol, for $1 \le i \le m$. Note that the outputs of the SM protocol are known 
only to $P_1$. After this, by applying homomorphic properties on $E_{pk}((x_i-y_i)^2)$, $P_1$ 
computes $E_{pk}(|X - Y|^2)$ locally based on Equation \ref{eq:euclidean}.
\begin{example} Refer to Table \ref{table:example}, let us assume that  
$P_1$ holds the encrypted data records of $t_1$ and $t_2$ as $X$ and $Y$ respectively. That is, 
$E_{pk}(X) = \langle E_{pk}(63), E_{pk}(1),$ $E_{pk}(1), E_{pk}(145), E_{pk}(233), 
E_{pk}(1), E_{pk}(3), E_{pk}(0), E_{pk}(6),$ $E_{pk}(0)\rangle$ and   
$E_{pk}(Y) = \langle E_{pk}(56), E_{pk}(1), E_{pk}(3), E_{pk}(130),$ $E_{pk}(256), 
E_{pk}(1), E_{pk}(2), E_{pk}(1), E_{pk}(6), E_{pk}(2)\rangle$. During the SSED protocol,  $P_1$ 
initially computes $E_{pk}(x_1-y_1) = E_{pk}(7), \ldots, E_{pk}(x_{10}-y_{10})= E_{pk}(-2)$. 
Then, $P_1$ and $P_2$ jointly compute $E_{pk}((x_1-y_1)^2) =E_{pk}(49) = SM(E_{pk}(7), E_{pk}(7)),\ldots, E_{pk}((x_{10}-y_{10})^2) 
= SM(E_{pk}(-2),$ $E_{pk}(-2)) = E_{pk}(4)$. $P_1$ locally computes $E_{pk}(|X - Y|^2) = 
E_{pk}(\sum_{i=1}^{10} (x_i - y_i)^2) = E_{pk}(813)$. 
%Similar, we can compute SSED for remaining records: 
%$E_{pk}(d_1)$ $ =E_{pk}(1549), E_{pk}(d_2) =E_{pk}(3622),E_{pk}(d_3) =E_{pk}(2080),$
%$E_{pk}(d_4)=E_{pk}(139),E_{pk}(d_6) =E_{pk}(12104)$.
\hfill $\Box$\\
\end{example}
\begin{algorithm}[!t]
\begin{algorithmic}[1]
\REQUIRE $P_1$ has $E_{pk}(X)$ and $E_{pk}(Y)$; $P_2$ has $sk$
\STATE $P_1$, \textbf{for} $1 \leq i \leq m$ \textbf{do}:
\begin{enumerate}\itemsep=0pt
    \item[(a).]  $E_{pk}({x_i-y_i}) \gets E_{pk}({x_i}) \ast E_{pk}(y_i)^{N-1}$   
\end{enumerate}
\STATE $P_1$ and $P_2$, \textbf{for} $1 \leq i \leq m$ \textbf{do}:
\begin{enumerate}\itemsep=0pt
    \item[(a).] Compute $E_{pk}((x_i-y_i)^2)$ using the SM protocol        
\end{enumerate}
\STATE $P_1$:
\begin{enumerate}\itemsep=0pt
    %\item[(a).] $\tau_i \gets E_{pk}({x_i-y_i})^2$, for $1 \le i \le m$        
    \item[(a).] $E_{pk}(|X - Y|^2) \gets \prod_{i=1}^m E_{pk}((x_i - y_i)^2)$
\end{enumerate}
\end{algorithmic}
\caption{SSED$(E_{pk}(X), E_{pk}(Y)) \rightarrow E_{pk}(|X - Y|^2)$ }
\label{alg:ssed}
\end{algorithm}
%%%%%%%%%%%%%%% Secure Bit-Decomposition (SBD)   %%%%%%%%%%%%%%%%%%%%%%%%%%%%%   
\noindent \textbf{Secure Bit-Decomposition (SBD). } 
We assume that $P_1$ has $E_{pk}(z)$ and $P_2$ has $sk$, where $z$ is not 
known to both parties and $0 \le z < 2^l$. The goal 
of the secure bit-decomposition (SBD)  
protocol is to compute the encryptions of 
the individual bits of binary representation of $z$\cite{schoenmaker-2006,bksam-asiaccs13}. That is, 
the output is $[z] =  \langle E_{pk}(z_1), \ldots, E_{pk}(z_l) \rangle$, 
where  $z_1$ and $z_l$ 
denote the most and least significant bits of $z$ respectively. At the end, the output $[z]$ is known 
only to $P_1$. 

Since the goal of this paper is not to investigate existing SBD protocols, we simply 
use the most efficient SBD protocol that was recently proposed in \cite{bksam-asiaccs13}.

\begin{example} Let us suppose that $z=55$ and $l=6$. Then the SBD protocol with private input $E_{pk}(55)$ gives 
$[55] = \langle E_{pk}(1), E_{pk}(1), E_{pk}(0),$ $E_{pk}(1), E_{pk}(1), E_{pk}(1)\rangle$ as the output to $P_1$.
\hfill $\Box$\\
\end{example} 
%%%%%%%%%%%%%%% Secure Minimum (SMIN)   %%%%%%%%%%%%%%%%%%%%%%%%%%%%%   
\noindent \textbf{Secure Minimum (SMIN). }
In this protocol, $P_1$ with 
input $([u], [v])$ and $P_2$ with $sk$ securely compute 
the encryptions of the individual bits of $\min(u, v)$, i.e., 
the output is $[\min(u, v)]$. Here $[u] = \langle E_{pk}(u_1), \ldots, E_{pk}(u_l) \rangle$ 
and $[v] = \langle E_{pk}(v_1), \ldots, E_{pk}(v_l) \rangle$, where $u_1$ (resp., $v_1$) and 
$u_l$ (resp., $v_l$) are the most and least significant bits of $u$ (resp., $v$). At 
the end, the output $[\min(u, v)]$ is known only to $P_1$. 

\begin{algorithm}[!t]
%\footnotesize
\begin{algorithmic}[1]
\REQUIRE $P_1$ has $[u]$ and $[v]$, where $0 \leq u,v < 2^l$; $P_2$ has $sk$\\
\STATE $P_1$:
\begin{enumerate}\itemsep=0pt
        \item[(a).] Randomly choose the functionality $F$
        \item[(b).]  \textbf{for} $i=1$ to $l$ \textbf{do}:
                  \begin{itemize}\itemsep=0pt
                           \item $E_{pk}(u_i*v_i) \gets \textrm{SM}(E_{pk}(u_i), E_{pk}(v_i))$  
                           \item $\textbf{if}~F : u > v ~\textbf{then}$:
                           \begin{itemize}\itemsep=0pt                              
                              \item $W_i \gets E_{pk}(u_i)\ast E_{pk}(u_i\ast v_i)^{N-1}$                               
                              \item $\Gamma_i \gets E_{pk}(v_i-u_i)\ast E_{pk}(\hat{r}_i)$; $\hat{r}_i \in_{R}\mathbb{Z}_{N}$
                           \end{itemize}
                               \textbf{else} 
                           \begin{itemize}\itemsep=0pt                               
                              \item $W_i \gets E_{pk}(v_i)\ast E_{pk}(u_i\ast v_i)^{N-1}$                                 
                              \item $\Gamma_i \gets E_{pk}(u_i-v_i)\ast E_{pk}(\hat{r}_i)$; $\hat{r}_i \in_{R}\mathbb{Z}_{N}$ 
                           \end{itemize}
                           \item $G_i \gets E_{pk}(u_i\oplus v_i)$
                           \item $H_i \gets H_{i-1}^{r_i} \ast G_i$; $r_i \in_{R}\mathbb{Z}_{N}$ and $H_0 = E_{pk}(0)$ 
                           \item $\Phi_i \gets E_{pk}(-1)\ast H_i$  
                           \item $L_i \gets W_i \ast \Phi_i^{r'_i}$; $r'_i \in_{R} \mathbb{Z}_{N}$  
                  \end{itemize}
    \item[(c).]  $\Gamma' \gets \pi_1(\Gamma)$  
    \item[(d).]  $L' \gets \pi_2(L)$; send $\Gamma'$ and $L'$ to $C$
\end{enumerate}
\STATE $P_2$:
\begin{enumerate}\itemsep=0pt
        \item[(a).] Receive $\Gamma'$ and $L'$ from $P_1$
        \item[(b).]  $M_i \gets D_{sk}(L'_i)$, for $1 \leq i \leq l$                   
        \item[(c).] $\textbf{if}~\exists~j~\textrm{such that}~M_j = 1 ~\textbf{then}$ $\alpha \gets 1$\\                            
                $\textbf{else}$ $\alpha \gets 0$                     
       \item[(d).] $M'_i \gets {\Gamma'_i}^{\alpha}$ , for $1 \leq i \leq l$                    
       \item[(e).] Send $M'$ and $E_{pk}(\alpha)$ to $P_1$
\end{enumerate}
\STATE $P_1$:
\begin{enumerate}\itemsep=0pt
        \item[(a).] Receive $M'$ and $E_{pk}(\alpha)$ from $P_2$
        \item[(b).] $\widetilde{M} \gets \pi_1^{-1}(M')$  
        \item[(c).] \textbf{for} $i=1$ to $l$ \textbf{do}:
                  \begin{itemize}\itemsep=0pt
                           \item $\lambda_i \gets \widetilde{M}_i\ast E_{pk}(\alpha)^{N - \hat{r}_i}$
                           \item $\textbf{if}~F : u > v ~\textbf{then}$ $E_{pk}(\min(u,v)_i) \gets E_{pk}(u_i)\ast \lambda_i$\\                           
                           \textbf{else} $E_{pk}(\min(u,v)_i) \gets E_{pk}(v_i)\ast \lambda_i$                                                      
                  \end{itemize}  
\end{enumerate}               
\end{algorithmic}
\caption{SMIN$([u], [v]) \rightarrow [\min(u,v)]$}
\label{alg:sm2n}
\end{algorithm}
We assume that $0 \le u,v < 2^l$ and propose a novel 
SMIN protocol. The basic idea of the proposed SMIN protocol is 
for $P_1$ to randomly choose the functionality $F$ (by flipping a coin), where 
$F$ is either $u > v$ or $v > u$, and to obliviously execute $F$ with 
$P_2$. Since $F$ is randomly chosen and known only to $P_1$, the output of 
the functionality $F$ is oblivious to $P_2$. Based on the output and chosen $F$, 
$P_1$ computes $[\min(u, v)]$ locally using homomorphic properties. 

The overall steps involved in the SMIN protocol are shown in 
Algorithm \ref{alg:sm2n}. To start with, $P_1$ initially 
chooses the functionality $F$ as either $u > v$ or $v > u$ 
randomly. Then, using the SM protocol, $P_1$ computes $E_{pk}(u_i\ast v_i)$ 
with the help of $P_2$, for $1 \le i \le l$. Now, depending on $F$, 
$P_1$ proceeds as follows, for $1 \leq i \leq l$:
\begin{itemize}
\item If $F: u > v$, compute
\begin{eqnarray*}
W_i &= &E_{pk}(u_i)\ast E_{pk}(u_i\ast v_i)^{N-1}\\
 & =& E_{pk}(u_i *(1 -  v_i)) \\
\Gamma_i &= & E_{pk}(v_i - u_i)\ast E_{pk}(\hat{r}_i)\\ 
 &=&  E_{pk}(v_i - u_i + \hat{r}_i)
\end{eqnarray*}
\item If $F: v > u$, compute:
\begin{eqnarray*}
W_i &= &E_{pk}(v_i)\ast E_{pk}(u_i\ast v_i)^{N-1}\\
 & =& E_{pk}(v_i \ast(1- u_i)) \\
\Gamma_i &= & E_{pk}(u_i - v_i)\ast E_{pk}(\hat{r}_i)\\ 
 &=&  E_{pk}(u_i - v_i + \hat{r}_i)
\end{eqnarray*}
where $\hat{r}_i$ is a random number in $\mathbb{Z}_N$
\item Observe that if $F: u >v$, then $W_i = E_{pk}(1)$ only if $u_i > v_i$, 
and $W_i = E_{pk}(0)$ otherwise. Similarly, when $F: v > u$, we have $W_i = E_{pk}(1)$ only 
if $v_i > u_i$, and $W_i = E_{pk}(0)$ otherwise. Also, depending of $F$, $\Gamma_i$ stores the encryption of 
randomized difference between $u_i$ and $v_i$ which will be used in later computations. 
\item Compute the encrypted bit-wise XOR 
between the bits $u_i$ and $v_i$ as $G_i = E_{pk}(u_i \oplus v_i)$  using the 
below formulation:
\begin{center}%\footnotesize
$G_i = E_{pk}(u_i)\ast E_{pk}(v_i)\ast E_{pk}(u_i\ast v_i)^{N-2}$
\end{center} In general, for any two given bits $o_1$ and $o_2$, we have $o_1\oplus o_2 = o_1 + o_2 -2(o_1\ast o_2)$

\item Compute an encrypted vector $H$ by preserving the first occurrence of $E_{pk}(1)$ (if there 
exists one) in $G$ by initializing $H_0 = E_{pk}(0)$. The  
rest of the entries of $H$ are computed as $H_i =  H_{i-1}^{r_i}\ast G_i$. 
We emphasize that at most one of the entry in $H$ is $E_{pk}(1)$ and the remaining 
entries are encryptions of either 0 or a random number. Also, if there exists an index 
$j$ such that $H_j = E_{pk}(1)$, then index $j$ is the first position (starting from the most significant bit) 
at which the corresponding bits of 
$u$ and $v$ differ.

\item Then, $P_1$ computes $\Phi_i = E_{pk}(-1) \ast H_i$. Note that ``$-1$'' is 
equivalent to ``$N-1$'' under $\mathbb{Z}_N$. From 
the above discussions, it is clear that $\Phi_i = E_{pk}(0)$ at most once since $H_i$ is 
equal to $E_{pk}(1)$ at most once. Also, if $\Phi_j = E_{pk}(0)$, then index $j$ is the position 
at which the bits of $u$ and $v$ differ first.
\item Compute an encrypted vector $L$ by combining $W$ and $\Phi$. Note that $W_i$ stores 
the result of $u_i > v_i$  or $v_i > u_i$ which depends on $F$ known only to $P_1$. 
Precisely, $P_1$ computes $L_i = W_i \ast \Phi_i^{r'_i}$, where $r'_i$ is a 
random number in $\mathbb{Z}_N$. The observation here is if $\exists$ an index $j$ such 
that $\Phi_j = E_{pk}(0)$, denoting 
the first flip in the bits of $u$ and $v$, then $W_j$ stores the 
corresponding desired information, i.e., whether $u_j > v_j$ or $v_j > u_j$ in encrypted form. 
\end{itemize}
\renewcommand{\tabcolsep}{.15cm}
\begin{table}[!t]
\centering
\caption{ $P_1$ chooses $F$ as $v>u$ where $u=55$ and $v=58$} % title of Table
%\centering % used for centering table
\begin{tabular}{ccccccccccccc}  
     \hline %inserts double horizontal lines
     $[u]$&\;$[v]$&\;$W_i$ &\;$\Gamma_i$&\;$G_i$&\;$H_i$&\;$\Phi_i$&\;$L_i$&\;$\Gamma_i$'&\;$L'_i$&\;$M_i$&\;$\lambda_i$&\;$\min_i$\\ [1ex] % inserts table
     \hline % inserts single horizontal line
    	1&1&0& $r$ & $0$& $0$  &$-1$ &$r$&$1+ r$&$r$&$r$&$0$ & 1\\ 
    	1&1&0& $r$ & $0$& $0$  &$-1$ &$r$&$r$&$r$&$r$&$0$ & 1\\  
    	0&1&1& $-1 + r$ & $1$&$1$  &$0$ &$1$ &$1 + r$&$r$&$r$&$-1$& 0\\   
    	1&0&0& $ 1 + r$ & $1$&$r$&$r$&$r$&$-1 + r$&$r$&$r$&$1$ & 1\\   
    	1&1&0& $r$ &$0$& $r$&$r$&$r$&$r$&$1$ &$1$&$0$ & 1\\  
    	1&0&0& $1 + r$ & $1$&$r$&$r$&$r$&$r$&$r$&$r$&$1$ & 1\\  % [1ex] adds  
    \hline %inserts single line
\end{tabular}%}
\begin{tablenotes}
      \small
      \item All column values are in encrypted form except $M_i$ column. Also, 
$r \in_R \mathbb{Z}_N$ is different for each row and column. 
%Also, $r$ is a random number in $\mathbb{Z}_N$ which is different for each row. 
    \end{tablenotes}

\label{table:SMIN-example} % is used to refer this table in the text
\end{table} 
After this, $P_1$ permutes the encrypted vectors $\Gamma$ and $L$ using two random permutation 
functions $\pi_1$ and $\pi_2$. Specifically, $P_1$ computes $\Gamma' = \pi_1(\Gamma)$ and 
$L' = \pi_2(L)$, and sends them to $P_2$. Upon receiving, $P_2$ decrypts $L'$ component-wise to get 
$M_i = D_{sk}(L'_i)$, for $1 \leq i \leq l$, 
and checks for index $j$ (decide the output of $F$). That is, if 
$M_j = 1$, then the output of $F$ is 1, and 0 otherwise. Let the output 
be $\alpha$. Note that since 
$F$ is not known to $P_2$, the output $\alpha$ is oblivious to $P_2$. In addition, $P_2$ computes a new encrypted vector $M'$ where 
$M'_i = {\Gamma'_i}^{\alpha}$, for $1 \leq i \leq l$, sends $M'$ and $E_{pk}(\alpha)$ to 
$P_1$. After receiving $M'$ and $E_{pk}(\alpha)$, $P_1$ computes the inverse permutation of $M'$ 
as $\widetilde{M} = \pi_1^{-1}(M')$. Then, $P_1$ performs the following homomorphic 
operations to compute the encryption of $i^{th}$ bit of $\min(u, v)$, i.e., 
$E_{pk}(\min(u,v)_i)$, for $1 \leq i \leq l$:
\begin{itemize}
\item Remove the randomness from $\widetilde{M}_i$ by 
computing $$\lambda_i = \widetilde{M}_i\ast E_{pk}(\alpha)^{N -\hat{r}_i}$$
\item If $F: u>v$, compute the $i^{th}$ encrypted bit of $\min(u, v)$ as 
$E_{pk}(\min(u,v)_i) = E_{pk}(u_i)\ast \lambda_i = E_{pk}(u_i + \alpha*(v_i - u_i))$. Otherwise, compute
$E_{pk}(\min(u, v)_i) = E_{pk}(v_i)\ast \lambda_i = E_{pk}(v_i + \alpha*(u_i - v_i))$.
\end{itemize}
In the SMIN protocol, one main observation (upon 
which we can also justify the correctness of the final output) is that 
if $F:u > v$, then $\min(u,v)_i = (1-\alpha)*u_i + \alpha*v_i$ always holds, for $1 \le i \le l$. 
Similarly, if $F: v>u$, then $\min(u,v)_i = \alpha*u_i + (1-\alpha)*v_i$ always holds.

\begin{example}
Consider that $u = 55$, $v = 58$, and $l=6$. 
Assume that $P_1$ holds $[55]=\langle E_{pk}(1),E_{pk}(1),E_{pk}(0) ,E_{pk}(1),$ $E_{pk}(1),E_{pk}(1)\rangle$ and
$[58]=\langle E_{pk}(1),E_{pk}(1),E_{pk}(1),E_{pk}(0),E_{pk}(1),E_{pk}(0)\rangle$. In addition, we assume that 
$P_1$'s random permutation functions are as given below. 
\begin{table}[h]
\centering
\renewcommand{\arraystretch}{1.5}
\begin{tabular}{l l l l l l l l}
$i$ &=& \quad 1 & \quad 2 & \quad 3 & \quad 4 & \quad 5 & \quad 6\\
& & \quad $\downarrow$ & \quad $\downarrow$ & \quad $\downarrow$ & \quad $\downarrow$ & \quad $\downarrow$ & \quad $\downarrow$ \\
$\pi_1(i)$~~&=& \quad 6 & \quad 5 & \quad 4 & \quad 3 & \quad 2 & \quad 1\\
$\pi_2(i)$~~&=& \quad 2 & \quad 1 & \quad 5 & \quad 6 & \quad 3 & \quad 4\\
\end{tabular}
\end{table}
Without loss of generality, suppose $P_1$ chooses the functionality $F : v > u$. Then, 
various intermediate results based on the SMIN protocol 
are as shown in Table \ref{table:SMIN-example}. Following from Table \ref{table:SMIN-example}, we observe that: 
\begin{itemize}
     \item At most one of the entry in $H$ is $E_{pk}(1)$ ($ = H_3$) and the remaining 
entries are encryptions of either 0 or a random number in $\mathbb{Z}_N$.
     \item  Index $j=3$ is the first position at which the corresponding bits of 
$u$ and $v$ differ.
      \item  $\Phi_3 = E_{pk}(0)$ since $H_3$ is 
equal to $E_{pk}(1)$. Also, since $M_5=1$, $P_2$ sets $\alpha$ to 1.
\end{itemize}
At the end, only $P_1$ knows $[\min(u,v)] = [u] = [55]$.
\hfill $\Box$\\
\end{example}
%%%%%%%%%%%%%%% Secure Minimum out of $k$ Numbers (SMIN$_k$)   %%%%%%%%%%%%%%%%%%%%%%%%%%%%%   
\noindent \textbf{Secure Minimum out of $n$ Numbers (SMIN$_n$). } 
Consider $P_1$ with private input $([d_1], \ldots, [d_n])$ and $P_2$ with $sk$, where 
$0 \le d_i < 2^l$ and $[d_i] = \langle E_{pk}(d_{i,1}), \ldots, E_{pk}(d_{i,l})\rangle$, 
for $1 \le i \le n$. The goal of the SMIN$_n$ protocol is 
to compute $[\min(d_1, \ldots, d_n)] = [d_{\min}]$ without revealing any information about $d_i$'s 
to $P_1$ and $P_2$. Here we construct a new SMIN$_n$ protocol by utilizing SMIN  as 
the building block. The proposed SMIN$_n$ protocol is an iterative approach and it computes the desired 
output in an hierarchical fashion. In each iteration, minimum between a pair of values is computed and are 
feeded as input to the next iteration. Therefore, generating a binary execution 
tree in a bottom-up fashion. At the end, only $P_1$ knows the final result $[d_{\min}]$.  
\begin{algorithm}[t]
\begin{algorithmic}[1]
\REQUIRE $P_1$ has $([d_1], \ldots, [d_n])$; $P_2$ has $sk$
\STATE $P_1$: 
\begin{enumerate}\itemsep=0pt
  \item[(a).] $[d'_i] \gets [d_i]$, for $1 \le i \le n$
  \item[(b).] $num \gets n$
\end{enumerate}    
\STATE $P_1$ and $P_2$, \textbf{for} $i=1$ to $\left \lceil \log_2 n \right \rceil$:
\begin{enumerate}\itemsep=0pt
    \item[(a).] \textbf{for} $1 \leq j \leq \left \lfloor \frac{num}{2} \right \rfloor$: 
              \begin{itemize}
               \item $\textbf{if}~i = 1 ~\textbf{then}$: 
                             \begin{itemize}\itemsep=2pt
                                 \item $[d'_{2j-1}] \gets$~SMIN$([d'_{2j -1}], [d'_{2j}])$
                                 \item $[d'_{2j}] \gets 0$  
                             \end{itemize} 
                      \textbf{else}
                            \begin{itemize}\itemsep=2pt
                                 \item $[d'_{2i(j-1)+1}] \gets$~SMIN$([d'_{2i(j-1)+1}], [d'_{2ij-1}])$
                                  \item $[d'_{2ij-1}] \gets 0$ 
                            \end{itemize} 
                                                                  
                 \end{itemize}               
    \item[(b).]  $num \gets  \left \lceil \frac{num}{2} \right \rceil$
\end{enumerate}
\STATE $P_1$:
\begin{enumerate}\itemsep=2pt
    \item[(a.)] $ [d_{\min}] \gets [d'_1]$
\end{enumerate}
\end{algorithmic}
\caption{SMIN$_n([d_1],\ldots, [d_n]) \rightarrow [d_{\min}]$}
\label{alg:smkn}
\end{algorithm}

The overall steps involved in the proposed SMIN$_n$ protocol are highlighted in 
Algorithm \ref{alg:smkn}. Initially, $P_1$ assigns $[d_i]$ to a temporary vector $[d'_i]$, for 
$1 \le i \le n$. Also, he/she creates a global variable $num$ and initialize it to 
$n$, where $num$ represents 
the number of (non-zero) vectors involved in each iteration. Since the 
SMIN$_n$ protocol executes in a binary tree hierarchy (bottom-up fashion), we have 
$\left \lceil \log_2 n \right \rceil$ iterations, and in each iteration, the number of vectors 
involved varies. In the first iteration (i.e., $i=1$), $P_1$  
with private input $([d'_{2j-1}], [d'_{2j}])$  and $P_2$ with $sk$ involve 
in the SMIN protocol, for 
$1 \leq j \leq \left \lfloor \frac{num}{2} \right \rfloor$. At the end of the 
first iteration, only $P_1$ knows 
$[\min(d'_{2j-1}, d'_{2j})]$ and nothing is revealed to $P_2$, for $1 \leq j \leq \left \lfloor \frac{num}{2} \right \rfloor$.
Also, $P_1$ stores the result $[\min(d'_{2j-1}, d'_{2j})]$ in $[d'_{2j-1}]$, 
updates $[d'_{2j}]$ to zero and $num$ to $\left \lceil \frac{num}{2} \right \rceil$. 

During the $i^{th}$ iteration, only the non-zero vectors are involved, for $2 \leq i \leq \left \lceil \log_2 n \right \rceil$. 
For example, during second iteration (i.e., $i=2$), only $[d'_1], [d'_3]$, and so on are involved. Note that 
in each iteration, the output is revealed only to $P_1$ and $num$ is updated to $\left \lceil \frac{num}{2} \right \rceil$. 
At the end of the SMIN$_n$ protocol, $P_1$ assigns the final encrypted binary vector of global 
minimum value, i.e., $[\min(d_1, \ldots, d_n)]$ which is stored in $[d'_1]$ to $[d_{\min}]$.
 \tikzset{edge from parent/.style=
     {draw, edge from parent path={(\tikzparentnode) -- (\tikzchildnode)}}}
\begin{figure}[!t]
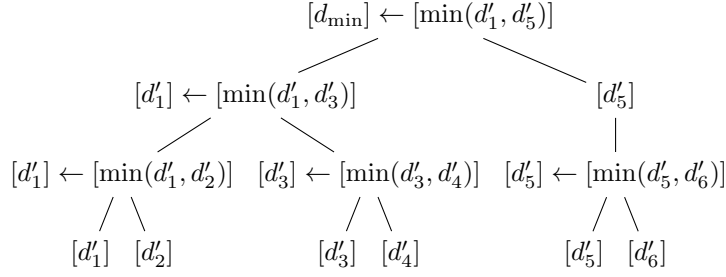

\centering
%\footnotesize \small

\Tree [.$[d_{\min}]\leftarrow [\min(d'_1,d'_5)]$ [.${[d'_1] \leftarrow [\min(d'_1,d'_3)]}$ [.${[d'_1] \leftarrow [\min(d'_1,d'_2)]}$ ${[d'_1]}$  ${[d'_2]}$ ] [.${[d'_3] \leftarrow[\min(d'_3,d'_4)]}$  ${[d'_3]}$ ${[d'_4]}$ ] ]
[.${[d'_5]}$ [.${[d'_5] \leftarrow [\min(d'_5,d'_6)]}$ ${[d'_5]}$ ${[d'_6]}$ ]  ] ]
\caption{Binary execution tree for $n=6$ based on the SMIN$_n$ protocol} \label{figure:SMIN_n-example}
%\vspace*{-0.6cm}
\end{figure}
 
For example, assume that $P_1$ holds $\langle [d_1], \ldots, [d_6]\rangle$ (i.e., $n=6$). 
Then, based on the SMIN$_n$ protocol, the binary execution tree (in a bottom-up fashion) 
to compute $[\min(d_1,\ldots, d_6)]$ is as shown in Figure \ref{figure:SMIN_n-example}. Note that, 
$[d'_i]$ is initially set to $[d_i]$, for $1 \le i \le 6$. \\\\
%%%%%%%%%%%%%%%  Secure Bit-OR (SBOR)   %%%%%%%%%%%%%%%%%%%%%%%%%%%%%   
\noindent \textbf{Secure Bit-OR (SBOR). }
Let us assume that $P_1$ holds $(E_{pk}(o_1), E_{pk}(o_2))$ and $P_2$ holds $sk$, where $o_1$ and $o_2$ are 
two bits not known to both parties. The goal of 
the SBOR protocol is to securely compute $E_{pk}(o_1 \vee o_2)$. At the end of this protocol, 
only $P_1$ knows $E_{pk}(o_1 \vee o_2)$. During this process, no information related to $o_1$ and 
$o_2$ is 
revealed to $P_1$ and $P_2$. Given the secure multiplication (SM) protocol, $P_1$ can compute 
$E_{pk}(o_1 \vee o_2)$ as follows:
\begin{itemize}
\item $P_1$ with input $(E_{pk}(o_1), E_{pk}(o_2))$ and $P_2$ with $sk$ involve in the SM protocol. At 
the end of this step, the output $E_{pk}(o_1*o_2)$ is known only to $P_1$. Note that, since 
$o_1$ and $o_2$ are bits, $E_{pk}(o_1*o_2) = E_{pk}(o_1 \wedge o_2)$.
\item $E_{pk}(o_1 \vee o_2) = E_{pk}(o_1 + o_2)\ast E_{pk}(o_1\wedge o_2)^{N-1}$.
\end{itemize} 
We emphasize that, for any given two bits $o_1$ and $o_2$, the property 
$o_1 \vee o_2 = o_1 + o_2 - o_1 \wedge o_2$ always holds. Note that, by 
homomorphic addition property, $E_{pk}(o_1 + o_2) = E_{pk}(o_1)\ast E_{pk}(o_2)$.
%Let $o_1=1$ and $o_1=0$, then  $1 \vee 0 = 1 + 0 - 1 \wedge 0 = 1$

%%%%%%%%%%%%%%%%%%%%%% proposed protocol %%%%%%%%%%%%%%%%%%%%%
\section{The Proposed Protocols}\label{sec:proposed}
In this section, we first present a basic S$k$NN protocol and demonstrate why such a simple 
solution is not secure. Then, we discuss our second approach, a fully secure 
$k$NN protocol. 
%present two different schemes: \textit{basic scheme} and textit{secure scheme} of secure $k$ nearest neighbor query problem. 
Both protocols are constructed using the security 
primitives discussed in Section \ref{sec:sub-methods} as building blocks. 

As mentioned earlier, we assume that Alice's 
database consists of $n$ records, denoted by $T = \langle t_1, \ldots, t_n \rangle$, 
and $m$ attributes, 
where $t_{i,j}$ denotes the $j^{th}$ attribute value of record $t_i$. Initially, Alice 
encrypts her database attribute-wise, that is, 
she computes $E_{pk}(t_{i,j})$, for $1 \le i \le n$ and $1 \le j \le m$. Let the encrypted database be denoted by $E_{pk}(T)$. We assume 
that Alice outsources $E_{pk}(T)$ as well as the future query processing service to the cloud. 
Without loss of generality, we assume that all attribute values and their Euclidean distances lie in $[0, 2^l)$. 

In our proposed protocols, we assume the existence of two non-colluding semi-honest 
cloud service providers, denoted by $C_1$ and $C_2$, which together form 
a federated cloud. We emphasize that such an assumption is not new and has been 
commonly used in the related problem domains\cite{twinclouds-2011,wang-fuzzy-2013}. The intuition behind 
such an assumption is as follows. Most of the cloud service providers in the market are well-established 
IT companies, such as Amazon and Google. Therefore,  a collusion 
between them is highly unlikely as it will damage their reputation which 
in turn effects their revenues. 

Under this setting, Alice outsources her encrypted database $E_{pk}(T)$ 
to $C_1$ and the secret key $sk$ to $C_2$. 
The goal of the proposed protocols is to retrieve the top 
$k$ records that are closest to the user query  
in an efficient and secure manner. Briefly, consider an authorized user Bob who wants 
to find $k$ records that are closest to his query 
record $Q = \langle q_1, \ldots, q_m\rangle $ based on $E_{pk}(T)$ in $C_1$. 
Bob initially sends his query $Q$ (in encrypted form) to $C_1$. After
this, $C_1$ and $C_2$ involve in a set of sub-protocols to securely retrieve 
(in encrypted form) the set of $k$ records corresponding 
to the $k$-nearest neighbors of the input query $Q$. At the end of 
our protocols, only Bob will receive the $k$-nearest neighbors to $Q$ as the output. 

\subsection{Basic Protocol}\label{sec:basic-method}

In the basic secure $k$-nearest neighbor query protocol, denoted by S$k$NN$_\textrm{b}$, 
we relax the desirable properties 
%in trade of the performance 
to produce an efficient protocol (more details are given in the later part of this section). 

The main steps involved in the \sknnb~protocol are given in Algorithm \ref{alg:basic}.
Bob initially encrypts his query $Q$ attribute-wise, that 
is, he computes $E_{pk}(Q) = \langle E_{pk}(q_1),\ldots, E_{pk}(q_m)\rangle$ and 
sends it to $C_1$. 
Upon receiving $E_{pk}(Q)$ from Bob, $C_1$ with private input $(E_{pk}(Q), E_{pk}(t_i))$ 
and $C_2$ with the secret key $sk$ jointly involve in the SSED protocol, where 
$E_{pk}(t_i) = \langle E_{pk}(t_{i,1}), \ldots, E_{pk}(t_{i,m})\rangle$, for $1 \le i \le n$. 
The output of this step, denoted by $E_{pk}(d_i)$, is the encryption of 
squared Euclidean distance between $Q$ and $t_i$, i.e., $d_i = |Q - t_i|^2$. As mentioned earlier, $E_{pk}(d_i)$ is 
known only to $C_1$, for $1 \le i \le n$. 
We emphasize that computation of exact Euclidean distance between encrypted 
vectors is hard to achieve as it involves square root. However, in our 
problem, it is sufficient to compare the squared Euclidean distances as 
it preserves relative ordering. After this, $C_1$ sends $\left\{\left\langle 1,E_{pk}(d_1)
\right\rangle,\dots,\left\langle n,E_{pk}(d_n)\right\rangle\right\}$ to $C_2$, where entry $\langle i, E_{pk}(d_i)\rangle$ 
correspond to data record $t_i$, for $1 \le i \le n$. 
Upon receiving $\left\langle 1,E_{pk}(d_1)\right\rangle,\dots,\left\langle n,E_{pk}(d_n)\right\rangle$, 
$C_2$ decrypts the encrypted distance in each entry to 
get $d_i = D_{sk}(E_{pk}(d_i))$. Then, $C_2$ generates an index list  
$\delta = \left\langle i_1,\ldots,i_k \right\rangle$  
such that $\left\langle d_{i_1},\ldots\,d_{i_k}\right\rangle$ are the top $k$ smallest distances 
among $\left\langle d_1,\dots,d_n \right\rangle$. After this, $C_2$ sends $\delta$ to $C_1$. 
Upon receiving $\delta$, $C_1$ proceeds as follows: 
\begin{itemize}
\item Select the encrypted records $E_{pk}(t_{i_1}),\dots,E_{pk}(t_{i_k})$ as the $k$-nearest 
records to $Q$ and randomize them attribute-wise. More specifically, $C_1$ computes 
$E_{pk}(\gamma_{j,h}) = E_{pk}(t_{i_j,h})* E_{pk}(r_{j,h})$, for $1 \le j \le k$ and $1\le h \le m$. 
Here $r_{j,h}$ is a random number in $\mathbb{Z}_N$ and $t_{i_j,h}$ denotes the column $h$ attribute 
value of data record $t_{i_j}$. Send $\gamma_{j,h}$ to $C_2$ and $r_{j,h}$ to Bob, for $1 \le j \le k$ and $1\le h \le m$. 
\end{itemize}
Upon receiving $\gamma_{j,h}$, for $1 \le j \le k$ and $1\le h \le m$, $C_2$ decrypts it 
to get $\gamma'_{j,h} = D_{sk}(\gamma_{j,h})$ and sends them to Bob. Note that, due to randomization 
by $C_1$, $\gamma'_{j,h}$ is always a random number in $\mathbb{Z}_N$.

Finally, upon receiving $r_{j,h}$ from $C_1$ and $\gamma'_{j,h}$ from $C_2$, 
Bob computes the attribute values of $j^{th}$ nearest neighbor to $Q$ as 
$t'_{j,h} = \gamma'_{j,h} - r_{j,h} \mod N$, for $1 \le j \le k$ and $1\le h \le m$. Note that 
$N$ is the RSA modulus or part of the public key $pk$.

%%%%%%%%%%%%%%%%%%% B A S I C   S C H E M E %%%%%%%%%%%%%%%%%%%%%%
\begin{algorithm}[!htbp]
\begin{algorithmic}[1]
\REQUIRE $C_1$ has $E_{pk}(T)$; $C_2$ has $sk$; Bob has $Q$
\STATE  Bob:
\begin{enumerate}\itemsep=0pt
     \item[(a).] Compute $E_{pk}(q_j)$, for $1 \le j \le m$     
     \item[(b).] Send $E_{pk}(Q)=\left\langle E_{pk}(q_1), \ldots, E_{pk}(q_m)\right\rangle$ to $C_1$
\end{enumerate}

\STATE $C_1$ and $C_2$:
\begin{enumerate}\itemsep=0pt
     \item[(a).] $C_1$ receives $E_{pk}(Q)$ from Bob
     \item[(b).] \textbf{for} $i=1$ to $n$ \textbf{do}: 
\begin{itemize}
      \item $E_{pk}(d_i) \gets \textrm{SSED}(E_{pk}(Q), E_{pk}(t_i))$
\end {itemize}     
\item[(c).] Send $\left\{\left\langle 1,E_{pk}(d_1)\right\rangle,\dots,\left\langle n,E_{pk}(d_n)\right\rangle\right\}$ to $C_2$
  \end{enumerate}

\STATE $C_2$:
\begin{enumerate}\itemsep=0pt
     \item[(a).] Receive $\left\{\left\langle 1,E_{pk}(d_1)\right\rangle,\dots,\left\langle n,E_{pk}(d_n)\right\rangle\right\}$ from $C_1$
     \item[(b).] $d_i \gets D_{sk}(E_{pk}(d_i))$, for $1 \le i \le n$
 %    \item[(b).] $\lambda \leftarrow$ {$D_{sk}(d_1),\dot,D_{sk}(d_n)$}% $for\; 1\leq i \leq n$
  %   \item[(c).] $\lambda' \leftarrow Sort(\lambda)$%, for\; 1\leq i \leq n$
     \item[(c).] Generate $\delta \gets \left\langle i_1,\dots,i_k \right\rangle$, such that 
$\left\langle d_{i_1},\ldots, d_{i_k}\right\rangle$ are the top $k$ smallest distances among $\left\langle d_1,\dots,d_n \right\rangle$
     \item[(d).] Send $\delta$ to $C_1$  
 \end{enumerate}

\STATE $C_1$:
\begin{enumerate}\itemsep=0pt
     \item[(a).] Receive $\delta$ from $C_2$
     \item[(b).] \textbf{for} $1 \le j \le k$ and $1 \le h \le m$ \textbf{do}:
       \begin{itemize}
                 \item $\gamma_{j,h} \gets E_{pk}(t_{i_j,h}) \ast E_{pk}(r_{j,h})$, where $r_{j,h} \in_R \mathbb{Z}_N$
                 \item Send $\gamma_{j,h}$ to $C_2$ and $r_{j,h}$ to Bob   
       \end{itemize}            
     %\item[(c).] Send $\langle  \gamma_1,\dots, \gamma_k \rangle$  to $C_2$
 \end{enumerate}

\STATE $C_2$:
\begin{enumerate}\itemsep=0pt
     %\item[(a).] Receive $\langle  \gamma_1,\dots, \gamma_k \rangle$ from $C_1$
     \item[(b).] \textbf{for} $1 \le j \le k$ and $1 \le h \le m$ \textbf{do}:
       \begin{itemize}
                \item Receive $\gamma_{j,h}$ from $C_1$
                \item $ \gamma'_{j,h} \gets D_{sk}(\gamma_{j,h})$; send $\gamma'_{j,h}$ to Bob 
       \end{itemize}           
 \end{enumerate}

\STATE  Bob:
\begin{enumerate}\itemsep=0pt
     \item[(a).] \textbf{for} $1 \le j \le k$ and $1 \le h \le m$ \textbf{do}:
       \begin{itemize}
                \item Receive $r_{j,h}$ from $C_1$ and $\gamma'_{j,h}$ from $C_2$     
                \item $t'_{j,h} \gets \gamma'_{j,h} - r_{j,h} \mod N$
       \end{itemize}           
\end{enumerate}

\end{algorithmic}
\caption{S$k$NN$_\textrm{b}(E_{pk}(T), Q) \rightarrow \langle t'_1,\dots,t'_k \rangle$}
\label{alg:basic}
\end{algorithm} 

%%%%%%%%%%%%%%%%%%%%%% proposed protocol %%%%%%%%%%%%%%%%%%%%%
\subsection{Fully Secure $k$NN Protocol}\label{sec:method}
The above-mentioned \sknnb~protocol reveals the data access patterns to $C_1$ and $C_2$. 
That is, for any given $Q$, $C_1$  and $C_2$ know which data records 
correspond to the $k$-nearest neighbors of $Q$. Also, 
it reveals $d_i$ values to $C_2$. However, leakage of such information may not be acceptable 
in privacy-sensitive applications such as medical data. Along this direction, 
we propose a fully secure protocol, denoted by \sknnm (where m stands for maximally secure), to retrieve 
the $k$-nearest neighbors of $Q$. The proposed \sknnm~protocol preserves all 
the desirable properties of a secure $k$NN protocol as mentioned in Section \ref{sec:intr}.

The main steps involved in the proposed \sknnm~protocol are as shown 
in Algorithm \ref{alg:main}. Initially, Bob sends his attribute-wise encrypted query $Q$, that 
is, $E_{pk}(Q) = \langle E_{pk}(q_1),\ldots, E_{pk}(q_m)\rangle$  
to $C_1$. Upon receiving, $C_1$ with private input $(E_{pk}(Q), E_{pk}(t_i))$ 
and $C_2$ with the secret key $sk$ jointly involve in the SSED protocol. The output 
of this step is $E_{pk}(d_i) = E_{pk}(|Q - t_i|^2)$ which will be known only to $C_1$, for 
$1 \le i \le n$. Then, $C_1$ with input $E_{pk}(d_i)$ and $C_2$ with $sk$ securely 
compute the encryptions of the individual bits of $d_i$ using the SBD protocol. Note 
that the output of this step $[d_i] = \langle E_{pk}(d_{i,1}), \ldots, E_{pk}(d_{i,l})\rangle$ is known 
only to $C_1$, where $d_{i,1}$ and $d_{i,l}$ are the most 
and least significant bits of $d_i$ respectively. Note that $0 \le d_i < 2^l$, for $1 \le i \le n$.

After this, $C_1$ and $C_2$ compute the top $k$ (in encrypted form) records that are 
closest to $Q$ in an iterative manner. More specifically, they compute 
$E_{pk}(t'_1)$ in the first iteration, $E_{pk}(t'_2)$ in the second iteration, and so on. Here $t'_s$ denotes 
the $s^{th}$ nearest neighbor to $Q$, for $1 \le s \le k$. At the end of 
$k$ iterations, only $C_1$ knows $\langle E_{pk}(t'_1), \ldots, E_{pk}(t'_k)\rangle$. 
To start with, in the first iteration, $C_1$ and $C_2$ jointly compute the encryptions 
of the individual bits of the minimum value among $d_1,\ldots, d_n$ using 
SMIN$_n$. That is, $C_1$ with input $\langle [d_1], \ldots, [d_n]\rangle$ 
and $C_2$ compute $[d_{\min}]$, where $d_{\min}$ is the minimum value 
among $d_1,\ldots, d_n$. The output $[d_{\min}]$ is known only to $C_1$. Now, $C_1$ performs 
the following operations locally:
\begin{itemize}
\item Compute the encryption of $d_{\min}$ from its encrypted individual bits as below 
\begin{eqnarray*}
E_{pk}(d_{\min}) &=& \prod_{\gamma=0}^{l-1} E_{pk}(d_{\min,\gamma+1})^{2^{l-\gamma -1}} \\
               &=& E_{pk}(d_{\min,1}\ast 2^{l-1} + \cdots + d_{\min, l})
\end{eqnarray*}
where $d_{\min,1}$ and $d_{\min,l}$ are the most and least significant bits of $d_{\min}$ respectively.
\item Compute the encryption of difference between $d_{\min}$ and each $d_i$. That is, 
$C_1$ computes $\tau_i = E_{pk}(d_{\min})\ast E_{pk}(d_i)^{N-1} = E_{pk}(d_{\min} - d_i)$, for $1 \le i \le n$. 
\item Randomize $\tau_i$ to get $\tau'_i = \tau_i^{r_i} = E_{pk}(r_i\ast(d_{\min} - d_i))$, where 
$r_i$ is a random number in $\mathbb{Z}_N$. Note that $\tau'_i$ is an encryption of 
either 0 or a random number, for $1 \le i \le n$. Also, permute $\tau'$ using a 
random permutation function $\pi$ (known only to $C_1$) to get $\beta = \pi(\tau')$ and 
send it to $C_2$.
\end{itemize} 
Upon receiving $\beta$, $C_2$ decrypts it component-wise to get $\beta'_i = D_{sk}(\beta_i)$, 
for $1 \le i \le n$. After this, he/she computes an encrypted vector $U$ of length $n$ 
such that $U_i = E_{pk}(1)$ if $\beta'_i=0$, and $E_{pk}(0)$ otherwise. Here we assume that 
exactly one of the entries in $\beta$ equals to zero and rest of them are random. This further 
implies that exactly one of the entries in $U$ is an encryption of 1 and 
rest of them are encryptions of 0's. However, we emphasize that if $\beta'$ has more than one 0's, 
then $C_2$ can randomly pick one of those indexes and 
assign $E_{pk}(1)$ to the corresponding index of $U$ and $E_{pk}(0)$ to the rest. Then, $C_2$ sends $U$ to $C_1$. 
After receiving $U$, $C_1$ performs inverse permutation on it to get 
$V = \pi^{-1}(U)$. Note that exactly one of the entry in $V$ is $E_{pk}(1)$ and the remaining are 
encryption of 0's. In addition, if $V_i = E_{pk}(1)$, then $t_i$ is the closest record 
to $Q$. However, $C_1$ and $C_2$ do not know which entry in $V$ corresponds 
to $E_{pk}(1)$. 

Finally, $C_1$ computes $E_{pk}(t'_1)$, encryption of the closest record 
to $Q$, and updates the distance vectors as follows: 
\begin{itemize}
\item $C_1$ and $C_2$ jointly involve in the secure multiplication (SM) protocol to compute 
$V'_{i,j} =  V_i \ast E_{pk}(t_{i,j})$, for $1 \le i \le n$ and $1 \le j \le m$.  The output $V'$ from the SM protocol is known only 
to $C_1$. After this, by using homomorphic properties, $C_1$ computes the 
encrypted record $E_{pk}(t'_1) = \langle E_{pk}(t_{1,1}), \ldots, E_{pk}(t_{1,m})\rangle$ locally, 
$E_{pk}(t'_{1,j})= \prod_{i=1}^n V'_{i,j}$, where $1 \le j \le m$. Note that $t'_{1,j}$ 
denotes the $j^{th}$ attribute value of record $t'_1$.
\item It is important to note that the first nearest tuple to $Q$ should be 
obliviously excluded from further computations. However, since $C_1$ does not 
know the record corresponding to $E_{pk}(t'_1)$, we need to obliviously eliminate 
the possibility of choosing this record again in next iterations. For this, 
$C_1$ obliviously updates the distance corresponding to $E_{pk}(t'_1)$  
to the maximum value, i.e., $2^l-1$. More specifically, $C_1$ updates the distance vectors 
with the help of $C_2$ using the SBOR protocol as below, for $1 \le i \le n$ and $1 \le \gamma \le l$.
\begin{center}$E_{pk}(d_{i,\gamma}) = \textrm{SBOR}(V_i, E_{pk}(d_{i,\gamma}))$\end{center}
Note that when $V_i = E_{pk}(1)$, the corresponding distance vector $d_i$ is set 
to the maximum value. That is, under this case, $[d_i] = \langle E_{pk}(1), \ldots, E_{pk}(1)\rangle $. 
However, when $V_i = E_{pk}(0)$, the OR operation has no affect on $d_i$. 
\end{itemize} 
The above process is repeated until $k$ iterations, and in each iteration $[d_i]$ corresponding 
to the current chosen record is set to the maximum value. However, since $C_1$ does 
not know which $[d_i]$ is updated, he/she has to re-compute $E_{pk}(d_i)$ in each iteration using 
their corresponding encrypted binary vectors $[d_i]$, for $1 \le i \le n$. 
In iteration $s$, $E_{pk}(t'_s)$ is known only to $C_1$.

At the end of the iterative step (i.e., step 3 of Algorithm \ref{alg:main}), 
$C_1$ has $\langle E_{pk}(t'_1), \ldots, E_{pk}(t'_k)\rangle$ - the list of encrypted 
records of $k$-nearest neighbors to the input query $Q$. The rest of the process is similar to 
steps 4 to 6 of Algorithm \ref{alg:basic}. Briefly, $C_1$ randomizes $E_{pk}(t'_j)$ attribute-wise 
to get $\gamma_{j,h} = E_{pk}(t'_{j,h})* E_{pk}(r_{j,h})$ and sends $\gamma_{j,h}$ to $C_2$ and $r_{j,h}$ to Bob, 
for $1 \le j \le k$ and $1 \le h \le m$. 
Here $r_{j,h}$ is a random number in $\mathbb{Z}_N$. Upon receiving $\gamma_{j,h}$'s, $C_2$ decrypts 
them to get the randomized $k$-nearest records as $\gamma'_{j,h} = 
D_{sk}(\gamma_{j,h})$ and sends them to Bob, for $1 \le j \le k$ and $1 \le h \le m$. 
Finally, upon receiving $r_{j,h}$ from $C_1$ and $\gamma'_{j,h}$ from $C_2$, 
Bob computes the $j^{th}$ nearest 
neighboring record to $Q$, as $t'_{j,h} = \gamma'_{j,h} - r_{j,h} \bmod N$, for $1 \le j \le k$ and $1 \le h \le m$.
\begin{algorithm}[!htbp]
%\footnotesize
\begin{algorithmic}[1]
\REQUIRE $C_1$ has $E_{pk}(T)$ and $\pi$; $C_2$ has $sk$; Bob has $Q$
\STATE  Bob sends $E_{pk}(Q) = \langle E_{pk}(q_1), \ldots, E_{pk}(q_m)\rangle $ to $C_1$
\STATE $C_1$ and $C_2$:
\begin{enumerate}\itemsep=-1pt
     %\item[(a).] $C_1$ receives $E_{pk}(Q)$ from Bob
     \item[(a).] \textbf{for} $i=1$ to $n$ \textbf{do:}
               \begin{itemize}
                      \item $E_{pk}(d_i) \gets \textrm{SSED}(E_{pk}(Q), E_{pk}(t_i))$ and $[d_i] \gets \textrm{SBD}(E_{pk}(d_i))$                      
               \end{itemize}         
\end{enumerate}
\STATE \textbf{for} $s=1$ to $k$ \textbf{do:}
\begin{enumerate}\itemsep=-1pt
    \item[(a).] $C_1$ and $C_2$:                            
              \begin{itemize}               
               \item $[d_{\min}] \gets$~SMIN$_n([d_1], \ldots, [d_n])$   
              \end{itemize}  
    \item[(b).] $C_1$:
              \begin{itemize}\itemsep=-1pt
                \item $E_{pk}(d_{\min}) \gets \prod_{\gamma=0}^{l-1} E_{pk}(d_{\min,\gamma+1})^{2^{l-\gamma -1}}$
                \item \textbf{if} $s \ne 1$ \textbf{then}, for $1 \le i \le n$
                \begin{itemize}       
                    \item $E_{pk}(d_i) \gets \prod_{\gamma=0}^{l-1} E_{pk}(d_{i,\gamma+1})^{2^{l -\gamma -1}}$  
                \end{itemize}      
                \item \textbf{for} $i=1$ to $n$ \textbf{do:}     
                \begin{itemize}\itemsep=0pt                                 
                    \item $\tau_i \gets E_{pk}(d_{\min})\ast E_{pk}(d_i)^{N-1}$
                    \item $\tau'_i \gets \tau_i^{r_i}$, where $r_i \in_R \mathbb{Z}_N$
                \end{itemize}  
                \item $\beta \gets \pi(\tau')$; send $\beta$ to $C_2$ 
              \end{itemize}  
    \item[(c).] $C_2$:
              \begin{itemize}\itemsep=-1pt
                %\item Receive $\beta$ from $C_1$
                \item $\beta'_i \gets D_{sk}(\beta_i)$, for $1 \le i \le n$
                \item Compute $U$, for $1 \le i \le n$:
                     \begin{itemize}\itemsep=-1pt   
                             \item \textbf{if} $\beta'_i = 0$ \textbf{then} $U_i = E_{pk}(1)$ 
                             \item \textbf{else} $U_i = E_{pk}(0)$
                     \end{itemize}
                \item Send $U$ to $C_1$     
               \end{itemize}
    \item[(d).] $C_1$:
              \begin{itemize}\itemsep=-1pt
                 %\item Receive $U$ from $C_2$
                 \item $V \gets \pi^{-1}(U)$
                 \item $V'_{i,j} \gets \textrm{SM}(V_i, E_{pk}(t_{i,j}))$, for $1 \le i \le n$ and $1 \le j \le m$  
                 \item $E_{pk}(t'_{s,j}) \gets  \prod_{i=1}^n V'_{i,j}$, for $1 \le j \le m$
                 \item  $E_{pk}(t'_s)= \langle E_{pk}(t'_{s,1}), \ldots, E_{pk}(t'_{s,m})\rangle$ 
              \end{itemize} 
    \item[(e).] $C_1$ and $C_2$, for $1 \le i \le n$:
              \begin{itemize}
                \item $E_{pk}(d_{i,\gamma}) \gets \textrm{SBOR}(V_i, E_{pk}(d_{i,\gamma}))$, for $1 \le \gamma \le l$
              \end{itemize}    
\end{enumerate}
\hspace{-0.45cm} The rest of the steps are similar to steps 4-6 of \sknnb
\begin{comment}
\STATE $C_1$:
\begin{enumerate}\itemsep=-1pt
     \item[(b).] Compute $ \gamma_1 \gets E_{pk}(t'_1) \ast E_{pk}(r),\dots, \gamma_k \gets E_{pk}(t'_k)\ast E_{pk}(r)$ where $r \in_R \mathbb{Z}_N$
     \item[(c).] Send $\langle  \gamma_1,\dots, \gamma_k \rangle$  to $C_2$ and $r$ to Bob
 \end{enumerate}

\STATE $C_2$:
\begin{enumerate}\itemsep=-1pt
     \item[(a).] Receive $\langle  \gamma_1,\dots, \gamma_k \rangle$ from $C_1$
     \item[(b).] Compute $ \gamma'_1 \gets D_{sk}(\gamma_1),\dots, \gamma'_k \gets D_{sk}(\gamma_k)$ 
     \item[(c).] Send $\langle \gamma'_1,\dots, \gamma'_k \rangle$ to Bob 
 \end{enumerate}

\STATE  Bob:
\begin{enumerate}\itemsep=-1pt
     \item[(a).] Receive $r$ from $C_1$ and $\langle \gamma'_1,\dots, \gamma'_k \rangle$ from $C_2$     
     \item[(b).] $t'_1 \gets \gamma'_1 - r,\dots,t'_k \gets \gamma'_k -r$
\end{enumerate}
\end{comment}
\end{algorithmic}
\caption{S$k$NN$_\textrm{m}(E_{pk}(T), Q) \rightarrow \langle t'_1, \ldots,t'_k\rangle$}
\label{alg:main}
\end{algorithm} 

%%%%%%%%%%%%%%%%%%%%%%%%%%%%%%%%%%%%%%%

\subsection{Security Analysis} 
Here we analyze the security guarantees of the proposed protocols. First, 
due to the encryption of $Q$ and by semantic security 
of the Paillier cryptosystem, Bob's input query $Q$ is protected from 
Alice, $C_1$ and $C_2$ in both protocols. 

In the \sknnb~protocol, 
the decryption operations at step 3(b) of Algorithm \ref{alg:basic} 
reveal $d_i$ values to $C_2$. In addition, since $C_2$ generates the top $k$ index list (at 
step 3(c) of Algorithm \ref{alg:basic}) and sends it to $C_1$, the data access patterns are revealed 
to $C_1$ and $C_2$. Therefore, our basic \sknnb~protocol is secure under 
the assumption that $d_i$ values can be revealed to $C_2$ and 
data access patterns can be revealed to $C_1$ and $C_2$. 

On the other hand, the security analysis of \sknnm~is as follows. At step 2 of Algorithm \ref{alg:main}, 
the outputs of SSED and SBD are in encrypted format, and are 
known only to $C_1$. In addition, all the intermediate results decrypted  
by $C_2$ in SSED are uniformly random in $\mathbb{Z}_{N}$. Also, 
as mentioned in \cite{bksam-asiaccs13}, the SBD protocol is secure. Thus, 
no information is revealed during step 2 of Algorithm \ref{alg:main}. In each 
iteration, the output of SMIN$_n$ is known only to $C_1$ and no information 
is revealed to $C_2$. Also, $C_1$ and $C_2$ do not know which record belongs 
to current global minimum. Thus, data access patterns are protected 
from both $C_1$ and $C_2$. At step 3(c) of Algorithm \ref{alg:main}, a component-wise decryption 
of $\beta$ reveals the tuples that satisfy the current global minimum 
distance to $C_2$. However, due to permutation by $C_1$, $C_2$ cannot trace back 
to the corresponding data records. Also, note that decryption 
of $\beta$ gives either encryptions of 0's or random numbers in $\mathbb{Z}_N$. 
Similarly, since $U$ is an encrypted 
vector, $C_1$ cannot know which tuple corresponds to current global minimum distance. 
Thus, data access patterns are further protected at this step from $C_1$. In addition, 
the update process at step 3(e) of Algorithm \ref{alg:main} does not leak any information 
to $C_1$ and $C_2$. In summary, $C_1$ and $C_2$ do not know which data records correspond  
to the output set $\langle t'_1, \ldots, t'_k\rangle$.

Based on the above discussions, it is clear that the proposed \sknnm~protocol 
protects the confidentiality of the data, privacy of user's input query, and hides 
the data access patterns.

\subsection{Complexity Analysis}\label{sec:complexity}
The computation complexity of \sknnb~is bounded by $O(n*m+k)$ encryptions, decryptions and 
exponentiations. In practice $k \ll n*m$; therefore, the computation complexity of 
\sknnb~is bounded by $O(n*m)$ encryptions and exponentiations (assuming that encryption and decryption 
operations under Paillier cryptosystem take similar amount of time). 

In the \sknnm~protocol, the computation complexity is bounded by $O(n)$ 
instantiations of SBD and SSED, $O(k)$ instantiations of SMIN$_n$, and $O(n*l)$ instantiations 
of SBOR. We emphasize that the computation complexity of the SBD protocol 
proposed in \cite{bksam-asiaccs13} is bounded by $O(l)$ encryptions and $O(l)$ exponentiations. 
%(under the assumption that encryption and decryption operations based on Paillier encryption scheme take 
%similar amount of time). 
Also, the computation complexity of SSED is bounded by $O(m)$ encryptions and $O(m)$ exponentiations. In 
addition, the computation complexity of SMIN$_n$ is bounded by $O(l\ast n\ast\log_2 n)$ encryptions 
and $O(l\ast n\ast \log_2 n)$ exponentiations. Since SBOR utilizes SM as a sub-routine, 
the computation cost of SBOR is bounded by (small) constant number of encryptions and exponentiations. 
Based on the above analysis, the total computation complexity of the \sknnm~protocol 
is bounded by $O(n\ast (l + m + k\ast l \ast \log_2 n))$ 
encryptions and exponentiations.

%%%%%%%%%%%%%%%%%%%%%%%%% Experiments %%%%%%%%%%%%%%%%%%%%%%%%
\section{Empirical Results} \label{sec:exp}
In this section, we discuss the  
performances of the proposed protocols in detail under different parameter settings. 
We used Paillier cryptosystem\cite{paillier-99} 
%as the underlying additive homomorphic encryption scheme 
and implemented the proposed protocols in C. Various experiments 
were conducted on a Linux machine with an 
Intel\textregistered~Xeon\textregistered~Six-Core\texttrademark~CPU 3.07 GHz 
processor and 12GB RAM running Ubuntu 10.04 LTS. 
%However, we also give some experimental results on Amazon EC2 in Section 5.3. 

Since it is difficult to control the parameters in a real dataset, 
we randomly generated synthetic datasets depending on the parameter values in consideration. 
Using these synthetic datasets we can  
perform a more elaborated analysis on the computation costs of the proposed protocols under different 
parameter settings. 
We encrypted these datasets attribute-wise, using the Paillier encryption 
whose key size is varied in our experiments, and the encrypted data were
stored on our machine. Based on the protocols protocols, we then 
executed a random query over this encrypted data. For the rest 
of this section, we do not discuss about the performance of Alice 
since it is a one-time cost. Instead, we evaluate 
and analyze the performances of \sknnb~and \sknnm~separately. In addition, 
we compare the two protocols. In all our experiments, the Paillier 
encryption key size, denoted by $K$, is set to either 512 or 1024 bits.

\subsection{Performance of \sknnb} 
In this sub-section, we analyze the computation costs of \sknnb~by 
varying the number of data records ($n$), number of 
attributes ($m$), number of nearest neighbors ($k$), and encryption key size ($K$). Note 
that \sknnb~is independent of the domain size of attributes ($l$). 

First, by fixing $k =5$ and $K=512$, we evaluated the computation costs 
of \sknnb~for varying $n$ and $m$. As shown in Figure \ref{fig:sknnb-512}, 
the computation costs of \sknnb~grows linearly with $n$ and $m$. For example, 
when $m=6$, the computation time of \sknnb~increases from 44.08 to 87.91 seconds 
when $n$ is varied from 2000 to 4000. A similar trend is observed for $K=1024$  
as shown in Figure \ref{fig:sknnb-1024}. For any fixed parameters, 
we observed that the computation time of \sknnb~increases almost by a factor 
of 7 when $K$ is doubled. 

Next, by fixing $m=6$ and $n=2000$, we evaluated the running times of \sknnb~for 
varying $k$ and $K$. The results are shown in Figure \ref{fig:sknnb-vary-K}. Irrespective 
of $K$, the computation time of \sknnb~does not change much with varying $k$. This is because 
most of the cost in \sknnb~comes from the SSED protocol which is independent of $k$. 
E.g., when $K=512$ bits, the computation time of \sknnb~changes from 44.08 to 44.14 seconds 
when $k$ is changed from 5 to 25. Based on the above discussions, it is clear that the running time of \sknnb~mainly 
depends on (or grows linearly with) $n$ and $m$ which further justifies our complexity 
analysis in Section \ref{sec:complexity}.

\subsection{Performance of \sknnm}
We also evaluated the computation costs of \sknnm~for 
varying values of $k$, $l$ and $K$. 
Throughout this sub-section, we fix $m=6$ and $n=2000$. However, we observed that the 
running time of \sknnm~grows linearly with $n$ and $m$.
\begin{figure*}[!tbp]
\centering
\subfigure[\sknnb~for $k=5$ and $K = 512$]
{
     \epsfig{file=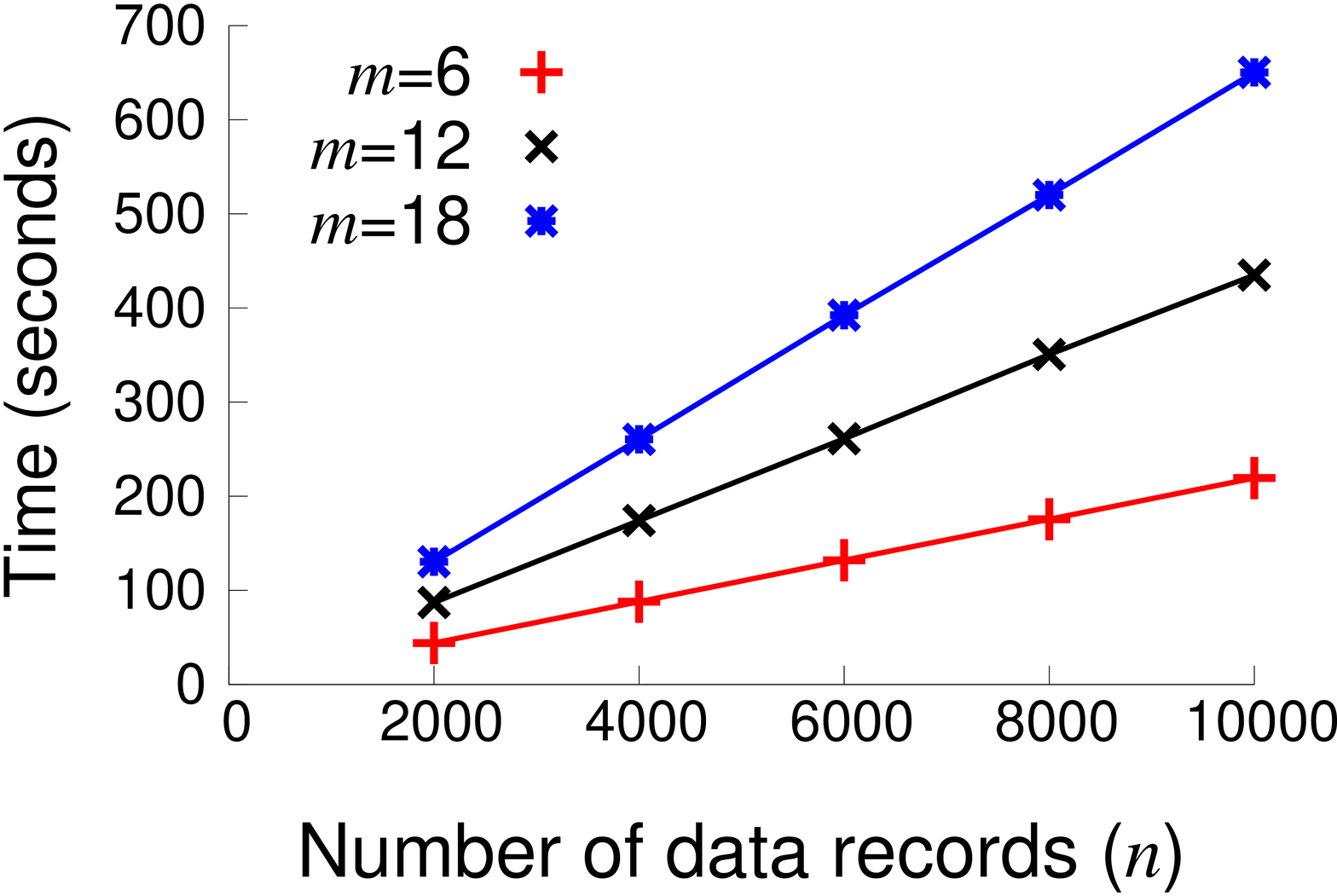, width= .31\textwidth}
\label{fig:sknnb-512}
}
\subfigure[\sknnb~for $k=5$ and $K = 1024$]
{
     \epsfig{file=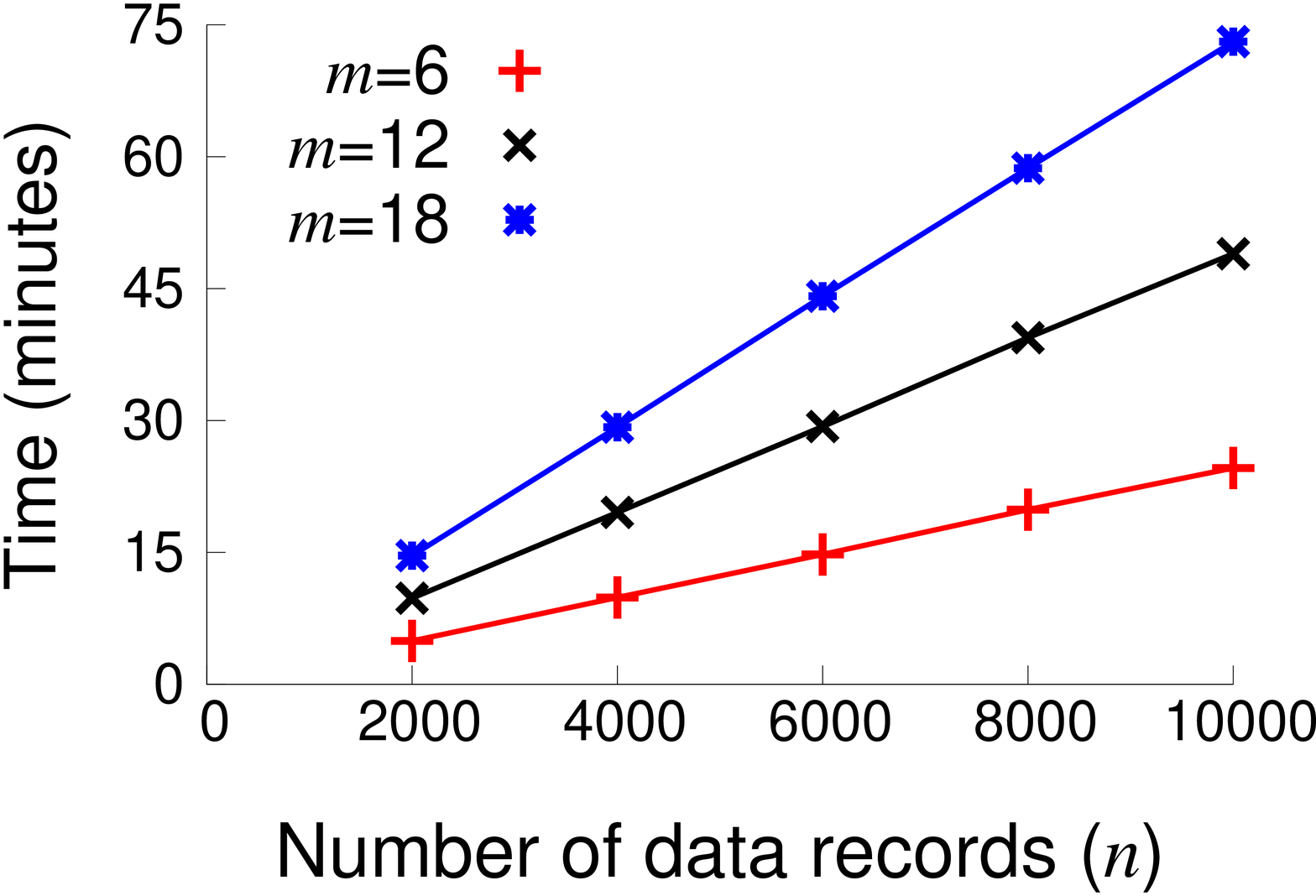, width= .31\textwidth}
\label{fig:sknnb-1024}
}
\subfigure[\sknnb~for $m=6$ and $n=2000$]
{
     \epsfig{file=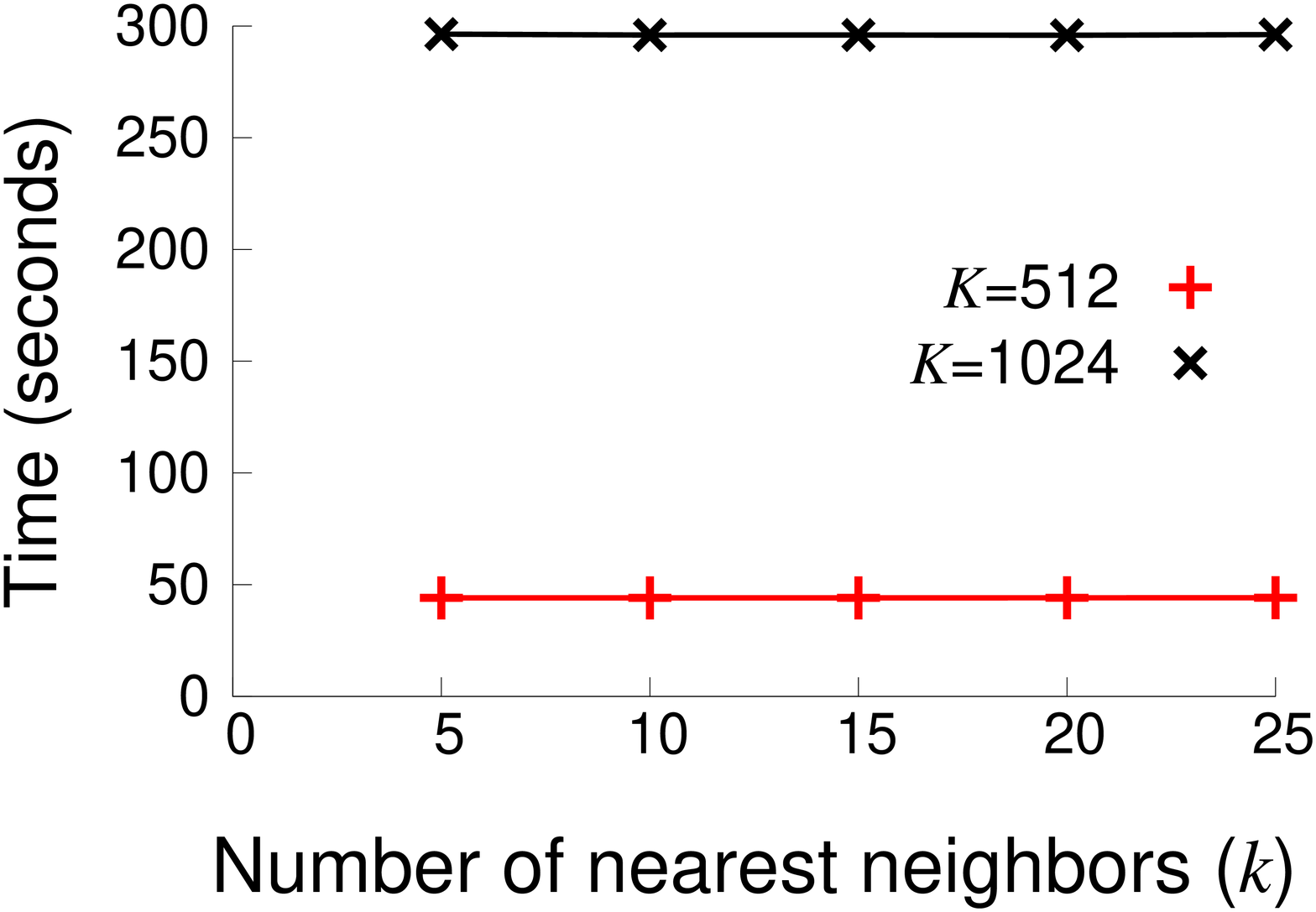, width= .31\textwidth}
\label{fig:sknnb-vary-K}
}
\subfigure[\sknnm~for $n=2000$ and $K = 512$]
{
     \epsfig{file=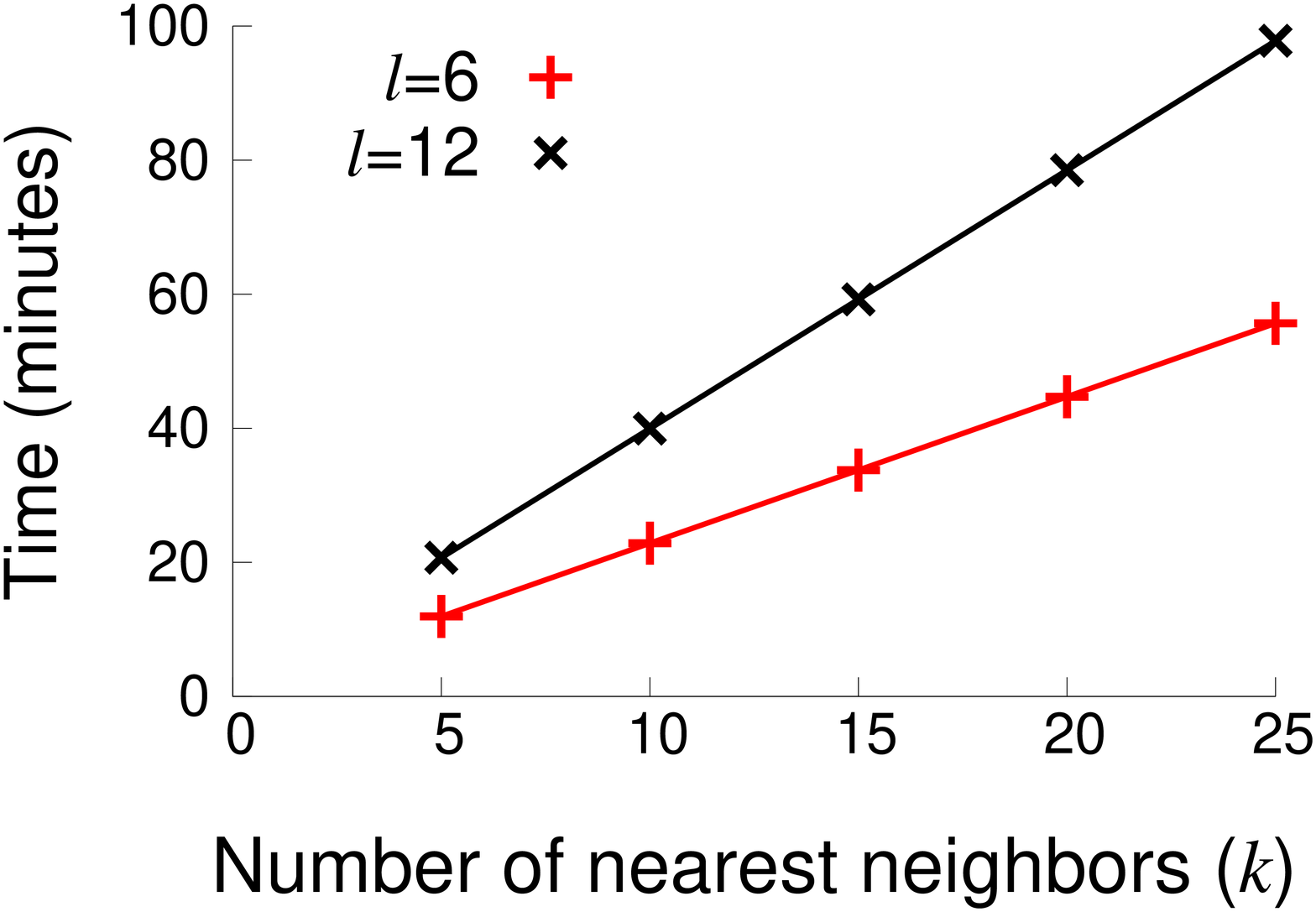, width= .31\textwidth}
\label{fig:sknnm-512}
}
\subfigure[\sknnm~for $n=2000$ and $K = 1024$]
{
     \epsfig{file=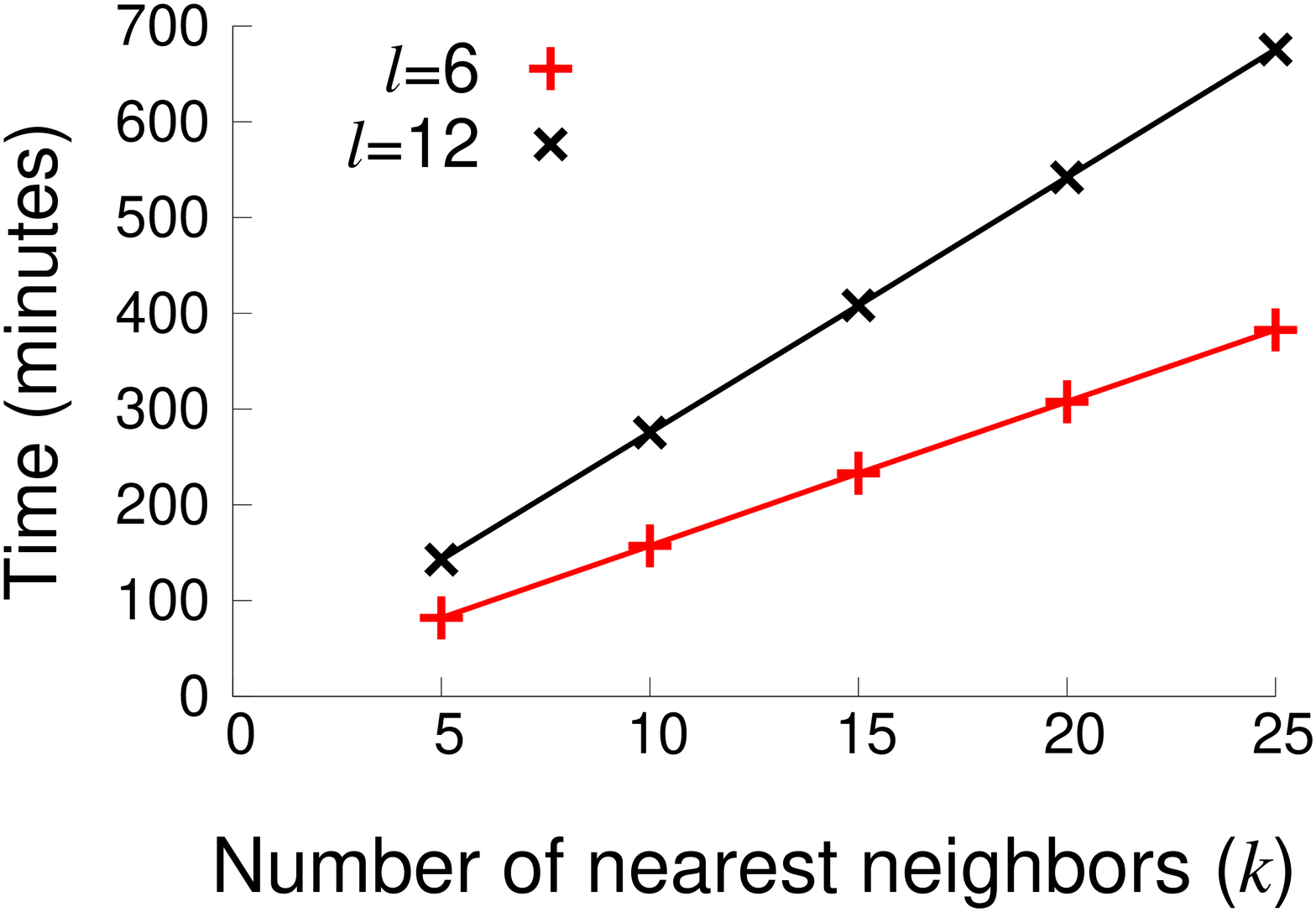, width= .31\textwidth}
\label{fig:sknnm-1024}
}
\subfigure[\sknnb~Vs.~\sknnm~for $n=2000, m=6, l=6$ and $K=512$]
{
     \epsfig{file=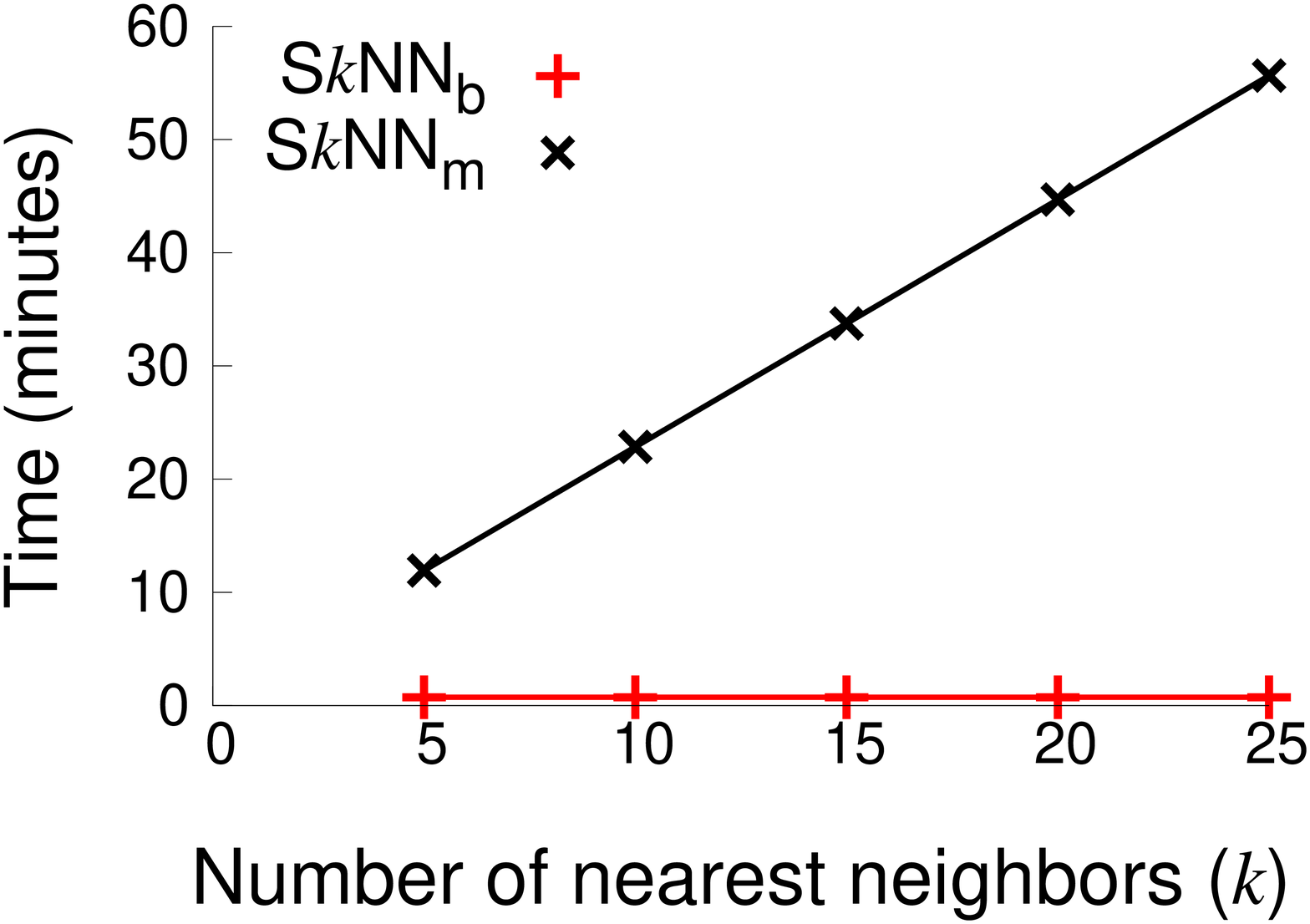, width= .31\textwidth}
\label{fig:comp-512}
}
\caption{Time complexities of \sknnb~and \sknnm~for varying values of $n$, $m$, $l$, $k$ and encryption key size $K$ }
\label{fig:complexity}
\end{figure*}

For $K$ = 512 bits, the computation costs of \sknnm~for varying $k$ and $l$ are 
as shown in Figure \ref{fig:sknnm-512}. Following from Figure \ref{fig:sknnm-512}, 
for $l=6$, the running time of \sknnm~varies from 11.93 to 55.65 minutes when $k$ is changed from 5 
to 25 respectively. Also, for $l=12$, the running time 
of \sknnm~varies from 20.68 to 97.8 minutes when $k$ is changed from 
5 to 25 respectively. In either case, the cost of \sknnm~grows 
almost linearly with $k$ and $l$. 

A similar trend is observed for $K=1024$ as shown in Figure \ref{fig:sknnm-1024}. In 
particular, for any given fixed parameters, we identified 
that the computation cost of \sknnm~increases by almost a factor 
of 7 when $K$ is doubled. For example, when $k$ = 10, \sknnm~ took 22.85 and  
157.17 minutes to generate the 10 nearest neighbors of $Q$ 
under $K$ = 512 and 1024 bits respectively. Furthermore, 
when $k$ = 5, we observed that around 69.7\% of cost in \sknnm~is accounted due to SMIN$_n$ which is 
initiated $k$ times in \sknnm~(once in each iteration). Also, the cost incurred 
due to SMIN$_n$ increases from 69.7\% to at least 75\% when $k$ is increased from 
5 to 25.

In addition, by fixing $n=2000, m=6, l=6$ and $K=512$, we compared the 
running times of both protocols for varying values of $k$. As shown in Figure \ref{fig:comp-512}, 
the running time of \sknnb~remains to be constant at 0.73 minutes since it 
is almost independent of $k$. However, 
the running time of \sknnm~changes from 11.93 to 55.65 minutes as we increase $k$ from 5 to 25. 

Based on the above results, it is clear that the computation costs 
of \sknnm~are significantly higher than that of \sknnb. However, we emphasize that \sknnm~is 
more secure than \sknnb; therefore, the two protocols act as a trade-off between security 
and efficiency. Also, it is important to note that Bob's computation cost is mainly due to the 
encryption of his input query record. As an example, for $m=6$, Bob's computation costs are 4 and 17 milliseconds 
when $K$ is 512 and 1024 bits respectively. This further shows that our proposed protocols are very efficient from 
end-user's perspective.

\subsection{Towards Performance Improvement}
\begin{figure}
\centering
\epsfig{file=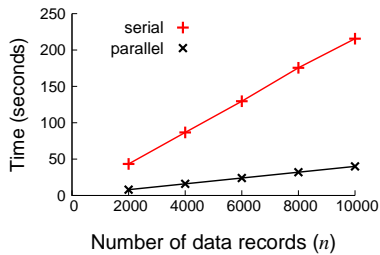, width= .31\textwidth}
\caption{Comparison of parallel and serial versions of \sknnb~for $m=6, k=5$ and $K=512$ bits}
\label{fig:comp-with-parallel}
\end{figure}

At first, it seems that the proposed protocols are costly and may 
not scale well for large datasets. However, in both protocols, we emphasize 
that the computations involved on each data record are 
independent of others. Therefore, we can parallelize the operations on data records for 
efficiency purpose. To 
further justify this claim, we implemented a 
parallel version of our \sknnb~protocol using OpenMP programming and compared its 
computation costs with its serial version. As mentioned earlier, our machine has 6 cores which can be used to 
perform parallel operations on 6 threads. For $m=6, k=5$ and $K=512$ bits, the comparison 
results are as shown in Figure \ref{fig:comp-with-parallel}. 
The observation is 
that the parallel version of \sknnb~is roughly 6 times more efficient than its serial version. This 
is because of the fact that the parallel version can execute operations on 6 data records at a time (i.e., 
on 6 threads in parallel). For example, when $n=10000$, the running times of 
parallel and serial versions of \sknnb~are 40 and 215.59 seconds respectively. 
 
%To further justify our claims, we considered an m2.4xlarge instance on 
%Amazon EC2\cite{amazon-ec2} that has 8 virtual cores which can be 
%used to perform parallel operations on 8 threads. We implemented a 
%parallel version of our \sknnb~protocol using MPI 
%programming and executed it on an m2.4xlarge instance on Amazon EC2 as the cloud. We 
%evaluated its efficiency gains over the non-parallel version of \sknnb. The observation is 
%that the parallel version of \sknnb~is roughly 8 times more efficient than its non-parallel version. This 
%is because of the fact that the parallel version executes operations on 8 data records at a time (i.e., 
%on 8 threads in parallel). 
%For example, when $n=1000, m=6$ and $K=512$, the running times of parallel and non-parallel 
%versions of \sknnb~are 21.8 and 3 seconds respectively. 

We believe that similar efficiency gains can be achieved by parallelizing the operations in \sknnm. 
%In particular to the SMIN$_n$ sub-routine in \sknnm, we can run the underlying SMIN sub-protocol in 
%parallel on each pair of entries. 
Based on the above discussions, especially in a cloud 
computing environment where high performance parallel processing can easily be 
achieved, we claim that the scalability issue of the proposed protocols can be eliminated or mitigated.
In addition, using the existing map-reduce techniques, we can drastically 
improve the performance further by executing parallel operations 
on multiple nodes. We will leave this analysis to future work. 

%The main advantages of the proposed PP$k$NN protocol are that it guarantees data confidentiality and privacy 
%of user's input query. In addition, it protects data access patterns and incurs 
%negligible computation costs on the end users.
%\balance
\section{Conclusion} \label{sec:concl}
Query processing is an important task in database management systems. In particular, 
$k$-nearest neighbors is one of the commonly used query in many data mining applications 
such as detection of fraud by credit card companies and prediction of tumor cells levels 
in blood. With the recent growth of cloud computing as a new IT paradigm, data owners are more 
interested to outsource their databases as well as DBMS functionalities to the cloud. Under an outsourced 
database environment, where encrypted data are stored in the cloud, secure query processing over 
encrypted data becomes challenging. To protect user privacy, various secure  
$k$-nearest neighbor (S$k$NN) techniques have been proposed in the literature. However, the existing 
S$k$NN techniques over encrypted data are not secure. 

Along this direction, we proposed 
two novel S$k$NN protocols over encrypted 
data in the cloud. The first protocol, which acts as a basic solution, leaks some information to 
the cloud. On the other hand, our second protocol is fully secure, that is, it protects the confidentiality of the data, user's 
input query, and also hides 
the data access patterns. However, the second protocol is more expensive compared to the basic protocol. 
Also, we evaluated the performance of our protocols under different parameter settings. As a future work, 
we will investigate and extend our research to other complex conjuctive queries over 
encrypted data.

\bibliographystyle{abbrv}
\bibliography{ref}

\begin{thebibliography}{10}

\bibitem{abadi2009data}
D.~J. Abadi.
\newblock Data management in the cloud: Limitations and opportunities.
\newblock {\em IEEE Data Eng. Bull}, 32(1):3--12, 2009.

\bibitem{agrawal2004order}
R.~Agrawal, J.~Kiernan, R.~Srikant, and Y.~Xu.
\newblock Order preserving encryption for numeric data.
\newblock In {\em Proceedings of the 2004 ACM SIGMOD international conference
  on Management of data}, pages 563--574. ACM, 2004.

\bibitem{twinclouds-2011}
S.~Bugiel, S.~N{\"u}rnberger, A.-R. Sadeghi, and T.~Schneider.
\newblock Twin clouds: An architecture for secure cloud computing (extended
  abstract).
\newblock In {\em Workshop on Cryptography and Security in Clouds (WCSC)},
  March 2011.

\bibitem{de2012managing}
S.~De~Capitani~di Vimercati, S.~Foresti, and P.~Samarati.
\newblock Managing and accessing data in the cloud: Privacy risks and
  approaches.
\newblock In {\em 2012 7th International Conference on Risk and Security of
  Internet and Systems (CRiSIS)}, pages 1--9. IEEE, 2012.

\bibitem{domingo2002provably}
J.~Domingo-Ferrer.
\newblock A provably secure additive and multiplicative privacy homomorphism.
\newblock {\em Information Security}, pages 471--483, 2002.

\bibitem{ghinita2008private}
G.~Ghinita, P.~Kalnis, A.~Khoshgozaran, C.~Shahabi, and K.-L. Tan.
\newblock Private queries in location based services: anonymizers are not
  necessary.
\newblock In {\em Proceedings of the 2008 ACM SIGMOD international conference
  on Management of data}, pages 121--132. ACM, 2008.

\bibitem{smc-2004}
O.~Goldreich.
\newblock {\em The Foundations of Cryptography}, volume~2, chapter General
  Cryptographic Protocols, pages 599--746.
\newblock Cambridge, University Press, Cambridge, England, 2004.

\bibitem{Goldreichnc}
O.~Goldreich.
\newblock {\em The Foundations of Cryptography}, volume~2, chapter Encryption
  Schemes, pages 373--470.
\newblock Cambridge University Press, Cambridge, England, 2004.

\bibitem{goldwasser-89}
S.~Goldwasser, S.~Micali, and C.~Rackoff.
\newblock The knowledge complexity of interactive proof systems.
\newblock {\em SIAM Journal of Computing}, 18:186--208, February 1989.

\bibitem{hacigumucs2004efficient}
H.~Hac{\i}g{\"u}m{\"u}{\c{s}}, B.~Iyer, and S.~Mehrotra.
\newblock Efficient execution of aggregation queries over encrypted relational
  databases.
\newblock In {\em Database Systems for Advanced Applications}, pages 125--136.
  Springer, 2004.

\bibitem{hore2012secure}
B.~Hore, S.~Mehrotra, M.~Canim, and M.~Kantarcioglu.
\newblock Secure multidimensional range queries over outsourced data.
\newblock {\em The VLDB Journal}, 21(3):333--358, 2012.

\bibitem{hore2004privacy}
B.~Hore, S.~Mehrotra, and G.~Tsudik.
\newblock A privacy-preserving index for range queries.
\newblock In {\em Proceedings of the Thirtieth international conference on Very
  large data bases-Volume 30}, pages 720--731. VLDB Endowment, 2004.

\bibitem{hu2011processing}
H.~Hu, J.~Xu, C.~Ren, and B.~Choi.
\newblock Processing private queries over untrusted data cloud through privacy
  homomorphism.
\newblock In {\em ICDE}, pages 601--612. IEEE, 2011.

\bibitem{uci-dataset-heart}
A.~Janosi, W.~Steinbrunn, M.~Pfisterer, and R.~Detrano.
\newblock Heart disease data set.
\newblock The UCI KDD Archive. University of California, Department of
  Information and Computer Science, Irvine, CA, 1988.
\newblock \url{http://archive.ics.uci.edu/ml/datasets/Heart+Disease}.

\bibitem{li2012toward}
M.~Li, S.~Yu, W.~Lou, and Y.~T. Hou.
\newblock Toward privacy-assured cloud data services with flexible search
  functionalities.
\newblock In {\em 32nd International Conference on Distributed Computing
  Systems Workshops (ICDCSW)}, pages 466--470. IEEE, 2012.

\bibitem{mell2011nist}
P.~Mell and T.~Grance.
\newblock The nist definition of cloud computing (draft).
\newblock {\em NIST special publication}, 800:145, 2011.

\bibitem{mykletun2006aggregation}
E.~Mykletun and G.~Tsudik.
\newblock Aggregation queries in the database-as-a-service model.
\newblock In {\em Data and Applications Security XX}, pages 89--103. Springer,
  2006.

\bibitem{paillier-99}
P.~Paillier.
\newblock Public-key cryptosystems based on composite degree residuosity
  classes.
\newblock In {\em EUROCRYPT}, Berlin, Heidelberg, 1999. Springer-Verlag.

\bibitem{pearson2009privacy}
S.~Pearson, Y.~Shen, and M.~Mowbray.
\newblock A privacy manager for cloud computing.
\newblock {\em Cloud Computing}, pages 90--106, 2009.

\bibitem{qi2008efficient}
Y.~Qi and M.~J. Atallah.
\newblock Efficient privacy-preserving k-nearest neighbor search.
\newblock In {\em the 28th International Conference on Distributed Computing
  Systems, 2008 (ICDCS'08)}, pages 311--319. IEEE, 2008.

\bibitem{bksam-asiaccs13}
B.~K. Samanthula and W.~Jiang.
\newblock An efficient and probabilistic secure bit-decomposition.
\newblock In {\em 8th ACM Symposium on Information, Computer and Communications
  Security (ASIACCS)}, pages 541--546, 2013.

\bibitem{schoenmaker-2006}
B.~Schoenmakers and P.~Tuyls.
\newblock Efficient binary conversion for paillier encrypted values.
\newblock In {\em EUROCRYPT}, pages 522--537. Springer-Verlag, 2006.

\bibitem{shaneck2009privacy}
M.~Shaneck, Y.~Kim, and V.~Kumar.
\newblock Privacy preserving nearest neighbor search.
\newblock {\em Machine Learning in Cyber Trust}, pages 247--276, 2009.

\bibitem{shi2007multi}
E.~Shi, J.~Bethencourt, T.-H. Chan, D.~Song, and A.~Perrig.
\newblock Multi-dimensional range query over encrypted data.
\newblock In {\em IEEE Symposium on Security and Privacy (SP'07)}, pages
  350--364. IEEE, 2007.

\bibitem{vaidya2005privacy}
J.~Vaidya and C.~Clifton.
\newblock Privacy-preserving top-k queries.
\newblock In {\em Proceedings of 21st International Conference on Data
  Engineering (ICDE '05)}, pages 545--546. IEEE, 2005.

\bibitem{wang-fuzzy-2013}
J.~Wang, H.~Ma, Q.~Tang, J.~Li, H.~Zhu, S.~Ma, and X.~Chen.
\newblock Efficient verifiable fuzzy keyword search over encrypted data in
  cloud computing.
\newblock {\em Computer Science and Information Systems}, 10(2):667--684, 2013.

\bibitem{williams2008building}
P.~Williams, R.~Sion, and B.~Carbunar.
\newblock Building castles out of mud: practical access pattern privacy and
  correctness on untrusted storage.
\newblock In {\em Proceedings of the 15th ACM conference on Computer and
  communications security}, CCS '08, pages 139--148. ACM, 2008.

\bibitem{wong2009secure}
W.~K. Wong, D.~W.-l. Cheung, B.~Kao, and N.~Mamoulis.
\newblock Secure knn computation on encrypted databases.
\newblock In {\em Proceedings of the 35th SIGMOD international conference on
  Management of data}, pages 139--152, 2009.

\bibitem{Yao82}
A.~C. Yao.
\newblock Protocols for secure computations.
\newblock In {\em Proceedings of the 23rd Annual Symposium on Foundations of
  Computer Science}, pages 160--164. IEEE Computer Society, 1982.

\bibitem{Yao86}
A.~C. Yao.
\newblock How to generate and exchange secrets.
\newblock In {\em Proceedings of the Symposium on Foundations of Computer
  Science}, pages 162--167. IEEE Computer Society, 1986.

\bibitem{yaosecure}
B.~Yao, F.~Li, and X.~Xiao.
\newblock Secure nearest neighbor revisited.
\newblock In {\em Proceedings of 29th IEEE International Conference on Data
  Engineering (ICDE)}, Brisbane, Australia, April 2013.

\end{thebibliography}

\end{document}